\documentclass[british]{zarticle}
\usepackage[export]{adjustbox}
\usepackage{slashed}
\usepackage[all]{xy}

\ifXeTeX
    \usepackage{xeCJK}
    \newcommand{\makeCJKtitle}{\maketitle}
\else
    \usepackage{CJK}
    \newcommand{\makeCJKtitle}{
        \begin{CJK*}{UTF8}{bkai}
            \maketitle
        \end{CJK*}}
\fi

\newcommand{\zhou}[1]{\textcolor{Blue}{\small [Zhou: \textit{#1}]}}
\newcommand{\yinchen}[1]{\textcolor{red}{\small [YCH: \textit{#1}]}}
\newcommand{\dg}[1]{\textcolor{green}{\small [DG: \textit{#1}]}}
\colorlet{Change}{BrickRed}

\newcommand{\cleancomment}{\renewcommand{\zhou}[1]{}\renewcommand{\yinchen}[1]{}\renewcommand{\dg}[1]{}\colorlet{Change}{black}}

\newcommand{\br}{\mathbf{r}}
\newcommand{\be}{\mathbf{e}}
\newcommand{\rd}{\mathrm{d}}
\newcommand{\rT}{\mathrm{T}}
\newcommand{\half}{\tfrac{1}{2}}
\newcommand{\halves}[1]{\tfrac{#1}{2}}
\newcommand{\hc}{\text{h.~c.}}
\newcommand{\CS}{\text{CS}}
\newcommand{\SU}{\mathrm{SU}}
\newcommand{\Sp}{\mathrm{Sp}}
\newcommand{\rU}{\mathrm{U}}
\newcommand{\SO}{\mathrm{SO}}
\newcommand{\rO}{\mathrm{O}}
\newcommand{\osp}{\mathfrak{osp}}
\newcommand{\OSp}{\mathrm{OSp}}
\newcommand{\CP}{\mathbb{C}\mathrm{P}}
\newcommand{\cL}{\mathcal{L}}

\newcommand{\cN}{\mathcal{N}}

\newcommand{\BZ}{\mathbb{Z}}
\newcommand{\BI}{\mathbb{I}}
\newcommand{\BR}{\mathbb{R}}
\newcommand{\bg}{\mathbf{g}}
\newcommand{\Yb}{\bar{Y}}
\DeclareMathOperator{\tr}{tr}

\DeclareMathOperator{\Pf}{Pf}

\begin{document}
\cleancomment

\title{Free and Interacting Fermionic Conformal Field Theories on the Fuzzy Sphere}
\date{\today}
\author[1,2]{Zheng Zhou (周正)}
\author[1]{Davide Gaiotto}
\author[3,1]{Yin-Chen He}
\affiliation[1]{Perimeter Institute for Theoretical Physics, Waterloo, Ontario N2L 2Y5, Canada}
\affiliation[2]{Department of Physics and Astronomy, University of Waterloo, Waterloo, Ontario N2L 3G1, Canada}
\affiliation[3]{C.~N.~Yang Institute for Theoretical Physics, Stony Brook University, Stony Brook, NY 11794-3840}

\abstract{The fuzzy-sphere regularisation is a powerful tool to study conformal field theories (CFT) in three spacetime dimensions. In this paper, we extend its scope to CFTs with local fermionic operators. We realise the free-Majorana-fermion CFT on a set-up with one flavour of bosons and one flavour of fermions on the lowest Landau level with a $1/2$ angular momentum mismatch and allow conversion between two bosons and two fermions, and use a relative chemical potential as the tuning parameter. On the phase diagram, we observe two continuous transitions described respectively by a free Majorana fermion and a gauged Ising CFT. We numerically confirm the emergent conformal symmetry through the operator spectrum and the two-point correlation function of the local Majorana fermion. We further establish a correspondence between the fuzzy-sphere models and the field-theory Lagrangians, and extend it to an interacting fermionic CFT --- the super-Ising theory with emergent super-conformal symmetry.}

\makeCJKtitle

\section{Introduction}
\label{sec:intro}

Conformal field theory (CFT) is one of the central topics of modern physics. CFT has provided important insights into various aspects of theoretical physics. In condensed-matter physics, it has produced useful predictions about the critical phenomena~\cite{Polyakov1970Conformal,Cardy1996Scaling,Sachdev2011Quantum}. Many classical and quantum phase transitions are conjectured to have emergent conformal symmetry in the infra-red (IR). In high-energy physics and quantum field theories, CFT also plays important role in understanding string theory~\cite{Polchinski1998String}, duality with gravitational theories on anti-de Sitter (AdS) space~\cite{Maldacena1998AdSCFT}, and renormalisation group flow structure~\cite{Zamolodchikov1986Irreversibility}. In 2D, the infinite-dimensional conformal algebra has made many theories exactly solvable~\cite{DiFrancesco1997CFT,Ginsparg1988CFT,Belavin1984BPZ}. However, going to the higher dimensions, the CFTs are less well-studied due to a much smaller conformal group, although some of the conformal data are determined at high precision by conformal bootstrap~\cite{Poland2018Bootstrap,Rychkov2023Bootstrap}.

Fermionic CFTs, \textit{i.~e.}~CFTs that include fermionic local operators, constitute a particularly interesting type of CFTs, including Gross-Neveu-Yukawa theories and certain gauge theories with or without Chern-Simons (CS) term. In some special cases, these CFTs can have emergent supersymmetry. For example, the theory of a single Majorana fermion $\chi$ coupled to a real Yukawa scalar field $\sigma$ flows to a super-conformal field theory (SCFT) with $\cN=1$ in the IR, known as the super-Ising theory~\cite{Fei:2016sgs,Grover2013SuperIsing,Rong2018SuperIsing,Atanasov2018SuperIsing,Atanasov2022SuperIsing}, where $\sigma$, $\chi$ and $\epsilon=\sigma^2$ combine into a super-conformal multiple and their scaling dimensions are locked together. Experimentally, these fermionic CFTs are easier to detect than their bosonic counterparts, as they produce rich electric signals, to which most experimental probes are naturally sensitive. For example, the frequency dependence of conductivity can be measured through electric transport experiments, and it corresponds to the OPE of the conserved symmetry current $J\times J$~\cite{Chowdhury2012JJCond,Katz2014JJCond}, including the central charge $C_J$ from its two-point function. An outstanding platform is the Moir\'e materials~\cite{Cai2023Moire,Zeng2023Moire,Park2023Moire}. Through tuning many available knobs like potential amplitude, they can be used to realise plateau transitions between fractional Chern insulators~\cite{Lee2018QEDCS,Song2023Moire,Song2023Moire2}. However, the nature and critical data of these theories remain poorly understood.

The fuzzy-sphere regularisation~\cite{Zhu2022} has emerged as a new powerful method to study 3D CFTs. This method involves studying quantum systems on a sphere that is `fuzzy' (non-commutative) due to a magnetic monopole at its centre~\cite{Ippoliti:2018ojo}. It offers distinct advantages including the exact preservation of rotation symmetry, the direct observation of emergent conformal symmetry and the efficient extraction of conformal data. In the fuzzy-sphere method, the state-operator correspondence plays an essential role. Specifically, there is a one-to-one correspondence between the eigenstates of the critical Hamiltonian on the sphere and the CFT operators, where the energy gaps are proportional to the scaling dimensions. The power of this approach has been first demonstrated in the context of the 3D Ising transition~\cite{Zhu2022}, where the presence of emergent conformal symmetry has been convincingly established. Since its proposal, various studies have greatly expanded its horizon, including
\begin{enumerate}
    \item accessing the `conformal data' of the 3D Ising CFT, such as the OPE coefficients~\cite{Hu2023Mar}, correlation functions~\cite{Han2023Jun}, entropic $F$-function~\cite{Hu2024}, conformal generators~\cite{Fardelli2024,Fan2024} and cross-cap coefficients~\cite{Dong2025},
    \item developing and applying techniques to improve the numerical precision, such as quantum Monte Carlo simulation~\cite{Hofmann2023} and conformal perturbation~\cite{Laeuchli2025},
    \item realising various 3D CFTs at integer filling, such as the free scalar~\cite{He2025Jun,Taylor2025} and Wilson-Fisher~\cite{Han2023Dec} CFTs, critical gauge theories realised through non-linear $\sigma$ model including $\SO(5)$~\cite{Zhou2023,Chen2024DQCP} and $\rO(4)$~\cite{Yang2025Jul} deconfined criticality and a series of new $\Sp(N)$ CFTs~\cite{Zhou2024Oct}, Potts model~\cite{Yang2025Jan}, and Yang-Lee criticality~\cite{Fan2025,ArguelloCruz2025,EliasMiro2025}, 
    \item exploring fractional quantum Hall (FQH) transitions, such as the Ising CFT on a FQH background~\cite{Voinea2024} and the confinement transition of the $\nu=1/2$ bosonic Laughlin state~\cite{Zhou2025Jul}, and
    \item studying the conformal defects (\textit{e.~g.}~the pinning-field defect~\cite{Hu2023Aug} and its $g$-function~\cite{Zhou2024Jan} and cusp~\cite{Cuomo2024}), conformal boundaries~\cite{Zhou2024Jul,Dedushenko2024} and lower-dimensional CFTs~\cite{Han2025}.
\end{enumerate}

All the CFTs that have been realised on fuzzy sphere so far have purely bosonic operator content. Apart from the fuzzy-sphere regularisation, a recent exploratory attempt to access the fermionic Gross-Neveu-Yukawa CFT considered Dirac fermions on a conventional sphere, using a basis of spinor spherical harmonics~\cite{Gao2025GNYSphere}. Although intuitive, this approach suffers from ultra-violet (UV) divergences similar to those in standard QFTs on continuous space, and thus further theoretical and numerical efforts may be required to justify its validity. In contrast, the fuzzy-sphere approach, similar to the highly successful lattice regularisation of QFTs and strongly interacting condensed-matter lattice models, is free of UV divergences~\cite{He2025Jun}.\footnote{In the continuum limit (\textit{i.~e.}~the monopole flux $q \to \infty$) of the fuzzy-sphere model, any (bare) local operator, as well as its correlators, remain non-diverging. Moreover, the microscopic couplings at the critical point also remain finite, \textit{i.~e.}, they converge to finite values in the continuum limit.} Consequently, the study of fermionic CFTs would greatly benefit from a fuzzy-sphere realisation, especially in light of the success achieved for their bosonic counterparts.

Nevertheless, the fuzzy-sphere realisation of fermionic CFTs faces some challenges. Most fuzzy-sphere models are built of microscopic fermions, \textit{i.~e.}~the fermions with $\rU(1)$ electric charge on the lowest Landau level. These microscopic fermions experience the magnetic field generated by the monopole and thus become fuzzy, in the sense that they have strong non-commutativity inherited from the magnetic field. For example, in flat space the microscopic fermions transform in a \emph{projective} representation of the continuous translation group, which is incompatible with any local operator of a CFT. The resolution is that these microscopic fermions must be gapped, so that the electric charge $\rU(1)$ is decoupled from the IR critical theory. Gapless `local' CFT operators on the fuzzy sphere tend to involve bilinear expressions in the microscopic fermions, which are charge neutral and bosonic. Moreover, another challenge is that the fermionic field in the CFT should carry half-integer spin (angular momentum), a requirement that microscopic fermions may not automatically satisfy. In this paper, we show that these challenges can be overcome by a fuzzy-sphere model with boson-fermion mixture. Concretely, we consider a set-up with both microscopic bosons $b$ and fermions $f$ with the same electric charge and let their angular momenta differ by $1/2$. Their bilinears $b^\dagger f$ or $f^\dagger b$ are neutral under electric charge and have fermionic statistics as well as half-integer spins. Therefore, the bilinears can be used to realise fermionic local operators in a CFTs.\footnote{A recent work~\cite{Zhou2025Jul} have also studied a boson-fermion mixture fuzzy-sphere model that realises the confinement transition of $\nu=1/2$ bosonic Laughlin state. Its set-up involves two flavours of charge-1 microscopic fermions and a flavour of charge-2 microscopic bosons. Hence, the charge-0 CFT operators are still bosonic.}

The simplest fermionic CFT to target is a massless free Majorana fermion. Its Lagrangian is
\begin{equation}
    \cL_{\text{Majorana}}=\frac{1}{2}\bar{\chi}(i\slashed{\partial}-m)\chi,
\end{equation}
where $\chi$ is a two component spinor, $\bar{\chi}=\chi^\rT\epsilon$, $\epsilon=i\gamma^2$,\footnote{In 3D Euclidean signature, the $\gamma$-matrices are identical to the Pauli matrices. We use $\gamma$ to avoid overloading the symbol $\sigma$.} and $\slashed{\partial}=\gamma^\mu\partial_\mu$. The free Majorana fermion $\chi$ has exact scaling dimension $\Delta_\chi=1$ and is subject to the equation of motion $\slashed{\partial}\chi=0$.\footnote{Unitarity and conformal symmetry imply that any local operator of spin-$\half$ with $\Delta=1$ is a free fermion, so this scaling dimension also gives a clear criterion to identify the free-fermion CFT.} It describes a transition driven by the mass $m$. Masses with different signs flow respectively to a trivially gapped phase and a $\nu_K=1$ Majorana quantum Hall (MQH) phase~\cite{Read1999pSC,Qi2010pSC}. The MQH is a gapped phase with no topological order characterised by thermal Hall conductance or chiral central charge $c_-=1/2$ and a chiral free Majorana fermion mode on the edge. The Kitaev index $\nu_K$ appears in the `16-fold way' classification of topological superconductors from spinless fermion~\cite{Kitaev2005Sixteen} and gives the number of chiral edge Majorana fermions and the chiral central charge $c_-=\nu_K/2$.\footnote{In this language, the $\nu=1$ fermionic integer quantum Hall phase can be regarded as a MQH phase with $\nu_K=2$.} It is also known as the $p$-wave topological chiral superconductor (TSC).\footnote{The name captures the parity and time-reversal breaking fermion pairing, although the TSC do not spontaneously break any $\rU(1)$ symmetry.}

In this paper, we study the set-up with one flavour of fermions and one flavour of bosons on the fuzzy sphere with the same electric charge but a mismatch $1/2$ in the angular momenta. We set the total filling $\nu=1$ and allow the conversion between two fermions and two bosons. By tuning a relative chemical potential, we identify a fermionic integer quantum Hall (fIQH) phase at $\nu=1$ when fermions are energetically favourable, a bosonic Pfaffian (bPf) phase~\cite{Read1999pSC,Moore1991Pfaffian} at $\nu_b=1$ with topological order characterised by a $\SU(2)_2$ Chern-Simons theory when bosons are energetically favourable, and an intermediate gapped phase corresponding to a $\nu_K=3$ MQH (\textit{i.~e.}~a $\nu_K=1$ MQH on the background of a $\nu=1$ fIQH state, or a $f$-wave TSC). We show theoretically and numerically that the fIQH-MQH transition is described by a free Majorana fermion. In particular, the energy gap, operator spectrum and two-point correlation function of local fermion all agree well with the free-Majorana-fermion CFT.

On the other hand, the MQH-bPf transition is described by an Ising CFT coupled to a $\BZ_2$ gauge field. The gauged Ising universality, also known as Ising$^\ast$, was first studied in the confinement transition of $\BZ_2$ topological order~\cite{Fradkin1978IsingGauge,Kogut1979IsingGauge,Jalabert1991IsingGauge,Sachdev1992IsingGauge,Senthil1999IsingGauge,Sachdev2000IsingGauge,Schuler2016IsingGauge}. While the Ising$^\ast$ criticality was first proposed to describe the transition between a trivially gapped state and a $\BZ_2$ topological order, in our set-up, confining a $\BZ_2$ gauge field drives the bPf phase to a MQH with trivial topological order. The only difference is a MQH background that does not change the universality. This is numerically confirmed by the agreement of the operators spectrum with the $\BZ_2$-even sector of the Ising CFT. An interesting feature of the gauged Ising CFT is the topological Wilson line defect. We show that a $\BZ_2$ gauge charge as the end-point of the Wilson line can be inserted by inserting a $2\pi$ monopole flux, and the end-point operators correspond to the $\BZ_2$-odd sector of the Ising CFT. 

We observe an analogy between the fuzzy-sphere models and the QFT Lagrangians in the realisation of both free Majorana fermion and critical real scalar. In particular, a flavour serves as the reference state from which the CFT ground state can be regarded as a perturabtion. The elementary fields that constitute the field theory are identified as hopping from the reference flavour to the excitation flavours. The interaction terms in the Hamiltonian can be identified with the terms in the QFT Lagrangian. We extend this correspondence to a simple interacting fermionic theory --- the super-Ising theory with a Majorana fermion coupled to a real scalar through a Yukawa interaction. We provide numerical evidence for the emergence of super-conformal symmetry and the agreement with the conformal bootstrap results.

In short, our boson-fermion mixture construction successfully extends the fuzzy-sphere approach to free and interacting fermionic CFTs, opening avenues to explore a broad class of theories, including the Gross-Neveu-Yukawa and quantum Hall transitions relevant to Moir\'e materials. A key ingredient is to create a half-integer fermionic field through the angular momentum offset of $1/2$ between microscopic fermions and bosons. This idea might be generalisable to define CFTs such that the most relevant operators have non-trivial spin: a `pure-symmetry' CFT such that the simplest local operators are conserved currents of spin-$1$, or a `pure-energy' CFT such that the simplest local operator is the stress tensor, a property expected from holographic duals to theories of quantum gravity~\cite{Witten2007PureEnerge,Maloney2007PureEnerge,Alday2022PureEnergy}. Both would be examples of `self-organised' critical theories stable to any deformation. It may also give a route to engineer spin-1 gauge fields and thus implement general renormalisable Lagrangians of quantum field theories on the fuzzy sphere. It may also be possible to reach effective field theories with higher spin quantum fields. The realisation of the gauged Ising CFT provides new possibilities in studying line defects like monodromy defect in Ising CFT. 

This paper is organised as the following: In Section \ref{sec:model}, we present the set-up of fuzzy sphere and our model Hamiltonian. In Section \ref{sec:phase}, we present the phase diagram supported by the numerical evidences for the gapped phases. In Section \ref{sec:critical}, we analyse the critical theories that describe the two continuous transitions between the three gapped phases. Then, Sections~\ref{sec:majorana} and \ref{sec:ising} are devoted to the numerical results respectively for the free-Majorana-fermion CFT and the gauged Ising CFT. In Section~\ref{sec:lag}, we establish the correspondence between fuzzy-sphere models and QFT Lagrangians and discuss the realisation of the super-Ising theory in light of this, and present in Section~\ref{sec:superising} the numerical evidence for the emergent super-conformal symmetry. Finally, in Section \ref{sec:discussion}, we make a brief summary and discuss the implications of our work and future directions.

\section{Model}
\label{sec:model}

In the set-up of the fuzzy sphere~\cite{Madore1991Fuzzy}, we consider one flavour of fermions $f$ and one flavour of bosons $b$ with the same electric charge on the sphere under the influence of a magnetic monopole at its centre. We let the monopole charge they experience differ by $2\pi$, \textit{i.~e.}, we let the fermion experience a $4\pi q$ monopole and the bosons experience a $4\pi(q-\half)$ monopole. As will be elaborated below, the mismatch of the monopole flux reflects the coupling between the electric charge and the spatial curvature, and it also allows the fermionic local operators like $f^\dagger b$ to have half-integer angular momenta. Due to the presence of the monopole, the single-particle eigenstates form highly degenerate quantised Landau levels~\cite{Haldane1983FQHE}. The lowest Landau level (LLL) has angular momentum $q$ and degeneracy $N_{mf}=2q+1$ for the fermions and angular momentum $q-\half$ and degeneracy $N_{mb}=2q$ for the bosons. By setting the single-particle gap the leading energy scale, we project the system onto the LLL and express the fermion and boson operators in terms of the creation and annihilation operators on the LLL
\begin{align}
    f^\dagger(\br)&=\frac{1}{R}\sum_{m=-q}^q f^\dagger_m Y_{qm}^{(q)}(\br),&b^\dagger(\br)&=\frac{1}{R}\sum_{m=-q+1/2}^{q-1/2}b^\dagger_mY_{q-1/2,m}^{(q-1/2)}(\br),
\end{align}
where the radius of the sphere is set as $R=N_{mf}^{1/2}$~\cite{Zhu2022}, and the single-particle wave-functions are the spin-weighted spherical harmonics~\cite{Wu1976LLL}.\footnote{An alternative choice without a flux mismatch is to let the bosons carry the two component spinor spherical harmonics $\Omega^{(q)}_{q-1/2,qm}$~\cite{Shnir:2005vvi} with spin-weight $q$ instead of the scalar spherical harmonics $Y^{(q-1/2)}_{q-1/2,m}$ with spin-weight $q-1/2$. These two choices assert the same spatial profile $\|\Omega^{(q)}_{q-1/2,qm}(\br)\|^2=|Y^{(q-1/2)}_{q-1/2,m}(\br)|^2$ and differ only by a local basis transformation of the spinors. The equivalence shows that the physical intepretation of the flux mismatch is the coupling of an additional spin-$1/2$ degree of freedom.}

The symmetries we preserve are:
\begin{enumerate}
    \item A $\rU(1)_e$ corresponding to the conservation of total electric charge $N_e=N_b+N_f$. This symmetry decouples from the low energy dynamics at the critical points, but it can still lead to non-trivial topological effect (\textit{cf.}~Section \ref{sec:ising}). We set $N_e=N_{mf}$ and allow the conversion between bosons and fermions.
    \item The $\SU(2)$ sphere rotation symmetry. The spatial parity is not manifest in our model, which frees us from the constraint of parity anomaly in theories like single free Majorana fermion. The conservation of fermion parity, whose charge is $(-1)^{N_f}$, is a part of this rotation symmetry. Because the boson and fermion orbitals have an angular momentum mismatch of $1/2$, the fermion parity even states have integer spin under $\SU(2)$ and the fermion parity odd states have half-integer spin under $\SU(2)$. In other words, we are building in spin-statistics rather than having it emerge in the relativistic IR CFT.
\end{enumerate}

The Hamiltonian we study includes the simplest symmetry-allowed terms:
\begin{enumerate}
    \item The local electric charge density interaction $H_U=\int\rd^2\br\,(n_f(\br)+n_b(\br))^2$, where $n_f(\br)=f^\dagger(\br)f(\br)$, $n_b(\br)=b^\dagger(\br)b(\br)$. This term is a natural form of interaction that exists in many other fuzzy-sphere models (\textit{e.~g.}~Refs.~\cite{Zhu2022,Zhou2023,He2025Jun,Zhou2025Jul}).
    \item A local boson-fermion pair conversion $H_t$ which annihilates two fermions and creates two bosons or vice versa. This term breaks the separate conservation of boson and fermion numbers to the total electric charge conservation. The conversion must occur in pair to preserve the fermion parity.
    \item A relative chemical potential $\mu N_f$, which enables us to energetically favour bosons or fermions.
\end{enumerate}
Altogether,
\begin{align}
    \label{eq:hmt}
    H&=H_U+tH_t+\mu N_f\\
    H_U&=\int\rd^2\br\,(n_f(\br)+n_b(\br))^2\nonumber\\
    &=\sum_{\{m_i\}}M_{m_1m_2m_3m_4}^{(b)}b^\dagger_{m_1}b^\dagger_{m_2}b_{m_3}b_{m_4}+M_{m_1m_2m_3m_4}^{(bf)}b^\dagger_{m_1}f^\dagger_{m_2}f_{m_3}b_{m_4}\nonumber\\
    H_t&=\sum_{\{m_i\}}M_{m_1m_2m_3m_4}^{(t)}b^\dagger_{m_1}b^\dagger_{m_2}f_{m_3}f_{m_4}+\hc\nonumber
\end{align}
The four-fermion contact terms are equivalently expressed in terms of the pseudo-potentials~\cite{Haldane1983FQHE,Zhu2022}
\begin{align*}
    M_{m_1m_2m_3m_4}^{(t)}&=\delta_{m_1+m_2,m_3+m_4}(4q-1)\begin{pmatrix}
       q-\half&q-\half & 2q-1\\ m_2&m_1 & -m_1-m_2
    \end{pmatrix}\begin{pmatrix}
        q&q &2q-1\\ m_3&m_4 & -m_3-m_4
    \end{pmatrix}\\
    M_{m_1m_2m_3m_4}^{(b)}&=\delta_{m_1+m_2,m_3+m_4}(4q-1)\begin{pmatrix}
        q-\half&q-\half & 2q-1 \\ m_2&m_1 & -m_1-m_2
    \end{pmatrix}\begin{pmatrix}
        q-\half&q-\half & 2q-1\\ m_3&m_4 & -m_3-m_4
    \end{pmatrix}\\
    M_{m_1m_2m_3m_4}^{(bf)}&=\delta_{m_1+m_2,m_3+m_4}\,4q\begin{pmatrix}
        q&q-\half & 2q-\half\\ m_2&m_1 & -m_1-m_2
    \end{pmatrix}\begin{pmatrix}
        q&q-\half & 2q-\half \\ m_3&m_4 & -m_3-m_4
    \end{pmatrix}.
\end{align*}
where $\left(\begin{smallmatrix}
    \bullet&\bullet&\bullet\\\bullet&\bullet&\bullet
\end{smallmatrix}\right)$ is the Wigner's $3j$-symbol. Note that in $H_t$, the orbital indices of the two bosons are symmetrised and fermions are anti-symmetrised. In real space, the pair conversion is equivalent to
\begin{align}
    H_t&=\int\rd^2\br\,\eta(\br)D_+\eta(\br)+\hc,&\eta(\br)&=b^\dagger(\br)f(\br),&D_\pm=D_\theta\pm iD_\phi
\end{align}
where $D_\mu$ is the covariant derivative on the sphere.\footnote{Specifically, $D^\mu=\partial^\mu+iqA^\mu$ and $A^\phi=\operatorname{ctg}\theta$. The derivatives act as the raising and lowering operators the spin-weighted spherical harmonics
\begin{equation*}
    D_\pm Y_{lm}^{(s)}=\pm\sqrt{(l\mp s)(l\pm s+1)}Y_{lm}^{(s\pm 1)}
\end{equation*}
The operator $\eta=b^\dagger f$ experiences a monopole flux $q=-1/2$.} This pair conversion can be understood as a kinetic term of the Majorana fermion (\textit{cf.}~Section \ref{sec:lag}).

Let us consider some limiting cases for $\mu$. At large negative $\mu$, fermions are energetically favourable and the fermion orbitals are fully filled, forming a fermionic integer quantum Hall (fIQH) state with $\nu=1$. At large positive $\mu$, bosons are energetically favourable, and the boson density interaction drives a bosonic Pfaffian (bPf) state~\cite{Moore1991Pfaffian,Regnault2004Pfaffian}. The mismatch of number of orbitals and particles $N_e=N_{mb}+1$ is in agreement with the Wen-Zee shift required by the bPf state~\cite{Wen1992Shift1,Wen1992Shift2} which arises from the coupling between the electric charge and the spatial curvature. We will discuss this phase in more detail in Secion~\ref{sec:phase}.

\section{Phase Diagram}
\label{sec:phase}

\begin{figure}[htbp]
    \centering
    \includegraphics[width=0.8\linewidth]{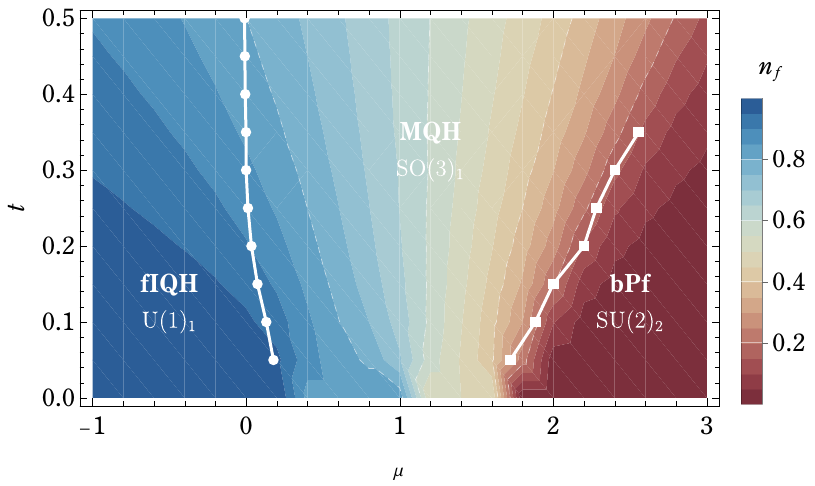}
    \caption{The phase diagram on the $\mu$-$t$ plane. The colour denotes the average fermion density calculated at $N_{mf}=12$. Selected contours are marked by dashed lines as visual guidance. The circles separate the $\nu=1$ fermionic integer quantum Hall (fIQH) phase and the $\nu_K=3$ Majorana quantum Hall (MQH) phase, and the squares separate MQH and bosonic Pfaffian (bPf) phase. The approximate phase boundary is determined through the scaling of the energy gap from $N_{mf}=12$ and $10$ data.}
    \label{fig:ph_diag}
\end{figure}

We study this model numerically through exact diagonalisation (ED) using our numerical package FuzzifiED~\cite{FuzzifiED}.\footnote{The module Fuzzifino, which is part of FuzzifiED, specialises in the ED for boson-fermion mixed systems. Sample codes can be found at \href{https://github.com/FuzzifiED/FuzzifiED.jl/blob/main/examples/free_majorana_spectrum.jl}{FuzzifiED.jl/examples/free\_majorana\_spectrum.jl}, \href{https://github.com/FuzzifiED/FuzzifiED.jl/blob/main/examples/free_majorana_correlator.jl}{free\_majorana\_correlator.jl}, and \href{https://github.com/FuzzifiED/FuzzifiED.jl/blob/main/examples/super_ising_spectrum.jl}{super\_ising\_spectrum.jl}.} The maximal system size we reach is $N_{mf}=13$, and the maximal RAM usage is 24 gigabytes. We first study the phase diagram on the $t$-$\mu$ plane. For finite $t$, we find that the fermion density drops continuously from $1$ to $0$ with increasing $\mu$ (Figure \ref{fig:ph_diag}).\footnote{For $t=0$, the number of fermions and bosons are separately conserved, level crossings appear for states in different $N_f$ sectors. The positive and negative $t$'s give exactly the same results, as they are connected by a $2\pi$-rotation along $z$ axis.}

\begin{figure}[htbp]
    \centering
    \includegraphics[width=0.49\linewidth]{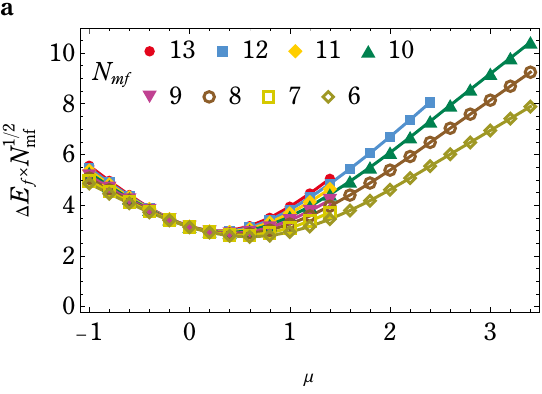}
    \includegraphics[width=0.49\linewidth]{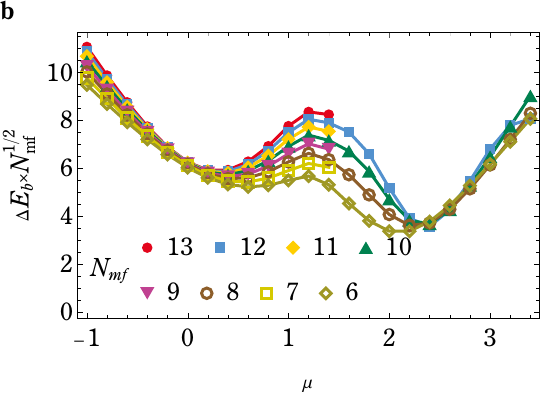}
    \caption{(a) The rescaled fermionic gap $\Delta E_f N_{mf}^{1/2}$ and (b) the rescaled bosonic gap $\Delta E_b N_{mf}^{1/2}$ as a function of $\mu$. In the calculation, we vary $N_{mf}$ and fix $t=0.3$. As bPf ground state only exists for even $N_{mf}$, we only keep the data with $\mu<1.5$ for odd $N_{mf}$.}
    \label{fig:gap}
\end{figure}

To study the nature of the fIQH-bPf transition, we first fix $t=0.3$. We calculate the finite-size behaviour of the fermionic gap $\Delta E_f$, defined as the energy difference between the lowest excited state with angular momentum $l=1/2$ and the ground state at $l=0$, and the bosonic gap $\Delta E_b$ defined as the energy difference between the two lowest $l=0$ states. At a Lorentz-invariant critical point, the gap is expected to close as $\Delta E\propto 1/R$. We find two such critical points. At $\mu_{c1}\approx0$, both bosonic and fermionic gap close; at $\mu_{c2}\approx2.3$, the bosonic gap closes while the fermionic gap remains (Figure \ref{fig:gap}). We keep the data only for even $N_{mf}$ near $\mu_{c2}$ because the bPf ground state is only compatible with even $N_{mf}$; for odd $N_{mf}$, the phase and the transition still exist, but the states have an anyon excitation (\textit{cf.}~Section \ref{sec:ising}). Similar scenario holds for other finite $t$. For each $t$, we determine the approximate phase boundary by comparing the energy gaps for $N_{mf,1}=12$ and $N_{mf,2}=10$ (Figure \ref{fig:ph_diag}): We minimise $[\Delta E(N_{mf,1})/\Delta E(N_{mf,2})](N_{mf,1}/N_{mf,2})^{1/2}$ respectively for the fermion gap and for the boson gap to get $\mu_{c1}$ and  $\mu_{c2}$. The minima are typically value close to $1$. 

The existence of two gapless points indicates that there is an intermediate phase whose transitions to fIQH and bPf phases are both continuous. To determine the nature of this intermediate phase, we calculate the real-space entanglement spectrum with the entanglement cut at the equator~\cite{Dubail2011RealEnt,Sterdyniak2011RealEnt}. For a 3D gapped topological phase, the low-lying part of the ground-state entanglement spectrum for a local bi-partition is generally believed~\cite{Li2008Edge} to reproduce the spectral counting of the physical edge CFT of that phase. In the following, we analyse the entanglement spectrum of the three phases respectively.

\begin{figure}[htbp]
    \centering
    \includegraphics[width=0.49\linewidth]{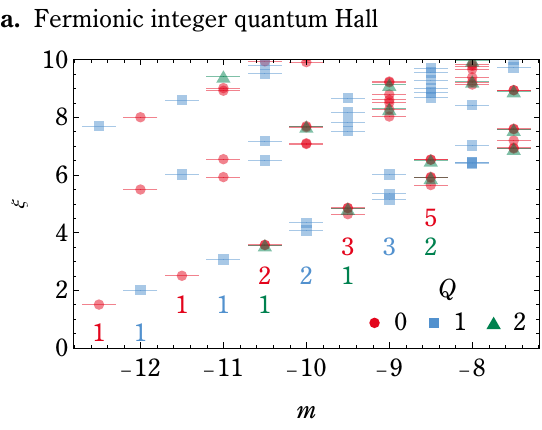}
    \includegraphics[width=0.49\linewidth]{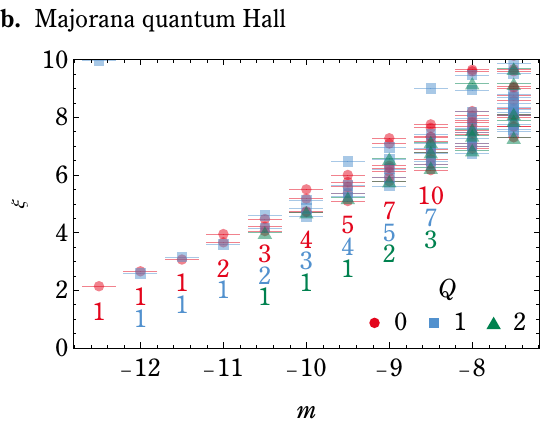}
    \includegraphics[width=0.49\linewidth]{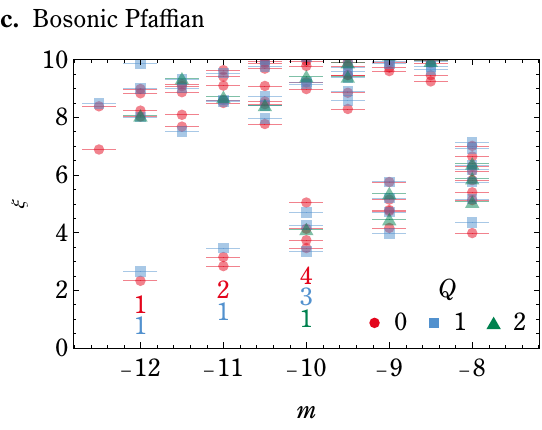}
    \caption{The real-space entanglement spectrum with the cut at the equator at different phases at different charge sector $Q=(Q_{e,A}-Q_{e,B})/2$: (a) the fermionic integer quantum Hall phase at $\mu=-2$, (b) the $\nu_K=3$ Majorana quantum Hall phase at $\mu=1$ and (c) the bosonic Pfaffian phase at $\mu=5$. Markers are plotted with transparency, so that nearly degenerate levels appear darker. The degeneracies for different charge sectors are labelled beneath. In the calculation we take $N_{mf}=10$.}
    \label{fig:ent}
\end{figure}

\paragraph{Fermionic integer quantum Hall phase (fIQH)} The large negative $\mu$ belongs to the $\nu=1$ fermionic integer quantum Hall phase. It can be described by a trivial Chern-Simons topological quantum field theory (TQFT) of $\rU(1)_1\cong\SO(2)_1$. The 2D chiral CFT on its edge is a free chiral Dirac fermion, or a pair of free chiral Majorana fermion. The two fermions form a $\rU(1)$ symmetry and carry charge $Q=\pm 1$. The $\rU(1)$ symmetry is manifest on the fuzzy sphere as the electric charge conservation. By state-operator correspondence, the degeneracy counts in the entanglement spectrum equal the numbers of operators with corresponding $\rU(1)$ charge $Q$ and angular momentum $m$ in the chiral 2D CFT. We give the detail for the calculation from the character in Appendix \ref{app:edge} (Eq.~\eqref{eq:edge_fiqh}) and label the expected degeneracies on Figure~\ref{fig:ent}. Practically $Q$ is the half the difference $(Q_{e,A}-Q_{e,B})/2$ between the charge in two hemispheres and $m$ is shifted by $-N_{mf}^2/8$. The counting is consistent with our numerical observation (Figure \ref{fig:ent}a).

\paragraph{Majorana quantum Hall phase (MQH)} We continue with the analysis of the intermediate phase. We shall show that this phase is a MQH phase with Kitaev index $\nu_K=3$~\cite{Kitaev2005Sixteen} (the $\nu=1$ fIQH can be viewed as $\nu_K=2$ MQH). The $\nu_K=3$ MQH is described by a $\SO(3)_1$ CS theory. The 2D chiral CFT on its edge is three free chiral Majorana fermions. They form a vector under a $\SO(3)$ global symmetry on the edge, of which a $\rU(1)$ subgroup is manifest on the fuzzy sphere as the charge conservation. In the entanglement spectrum, we observe a well-split lowest chiral edge mode whose degeneracy is consistent with the 2D chiral CFT (Figure \ref{fig:ent}b, Eq.~\eqref{eq:edge_mqh}). The TQFT has a chiral central charge $c_-=3/2$. The enhancement from $\rU(1)_e$ to the $\SO(3)$ global symmetry on the edge can be seen from the formation of multiplets. The excited states at $m=1/2$ and $1$ form $\SO(3)$-vectors corresponding to the Majorana fermion operator and the symmetry current. An intuition for its origin is the the fluctuating number of fermions in this phase and the pairing pattern given by $H_t$ which anti-symmetrises the orbital index of two fermions.

\paragraph{Bosonic Pfaffian (bPf)} The large positive $\mu$ belongs to the bosonic Pfaffian phase. The bosonic Pfaffian state is given by the many-body wave-function on a plane with complex co-ordinates $z,\bar{z}$~\cite{Moore1991Pfaffian}
\begin{equation}
    \Psi(\{z_i\})=\Pf\left(\frac{1}{z_i-z_j}\right)\prod_{i<j}(z_i-z_j)
\end{equation}
where $\Pf$ is the Pfaffian of an even-dimensional square matrix, which also restricts that the bPf state is only compatible with even number of particles, \textit{i.~e.} $N_{mf}\in 2\BZ$. It is described by a Chern-Simons theory $\SU(2)_2$. As $\SU(2)$ is a double covering of $\SO(3)$, $\SO(3)_1\cong\SU(2)_2/\BZ_2$, \textit{i.~e.}, the $\SU(2)_2$ can be regarded as gauging the $\BZ_2$ fermion-parity symmetry from $\SO(3)_1$. Correspondingly, its chiral edge mode is the same as removing the $m\in\BZ+\half$ levels and keeping the $m\in\BZ$ part of the chiral edge mode in $\SO(3)_1$ (Figure \ref{fig:ent}c). 

The $\SU(2)_2$ CS theory is a non-Abelian TO with chiral central charge $c_-^{\text{(bPf)}}=3/2$ and three kinds of anyons, \textit{viz.}~$\BI$, $\sigma$ and $\psi$. They have statistical angles and fusion rules
\begin{align}
    \theta^{\text{(bPf)}}_{\BI}&=0,&\theta^{\text{(bPf)}}_\psi&=\pi,&\theta^{\text{(bPf)}}_\sigma&=3\pi/8\nonumber\\
    \psi\times\psi&=\BI,&\psi\times\sigma&=\sigma,&\sigma\times\sigma&=\BI+\psi.
\end{align}
Here $\BI$ is the vacuum, $\psi$ is a fermionic anyon and $\sigma$ is a non-Abelian anyon. The anyons are also referred to as $0$, $\half$ and $1$. This TO can also be understood as an Ising topological order stacked with an Abelian topological order
\begin{equation*}
    \frac{\text{Ising}\times\rU(1)_4}{\BZ_2}.
\end{equation*}
The stacking with $\rU(1)_4$ adds $1$ to the chiral central charge and add an additional phase $\pi/4$ to the $\sigma$ anyon compared with their values $c_-^{(\text{Ising})}=1/2$ and $\theta^{\text{(Ising)}}_\sigma=\pi/8$ in the Ising TO. In Kitaev's `16-fold way' classification~\cite{Kitaev2005Sixteen}, this phase is described by the gauged MQH with index $\nu_K=3$.

\section{Critical Theories}
\label{sec:critical}

In this section, we discuss the critical theories that describe the transition between the fIQH, MQH, and bPf phases.

The transition from the fIQH to the MQH phase is described by a single free Majorana fermion. One way to see this is that the chiral central charge $c_-$ increases by $1/2$ across the transition. This transition is captured by the Lagrangian
\begin{equation}
    \cL_{\text{Majorana}}=\frac{1}{2}\bar{\chi}(i\slashed{\partial}-m)\chi+\frac{5}{4}\CS_g
\end{equation}
where $\chi$ is a (two-component) Majorana fermion, $\bar{\chi}=\chi^\rT\epsilon$, $\epsilon=i\gamma^2$, and $\slashed{\partial}=\gamma^\mu\partial_\mu$. The gravitational Chern-Simons term is defined through extending into a 4D manifold $X$ with the original 3D manifold its boundary $M=\partial X$~\cite{Hsin2016LevelRank}
\begin{equation*}
    \int_{M=\partial X}\CS_g=\frac{1}{192\pi}\int_X\tr R\wedge R
\end{equation*}
where $R$ is the curvature 2-form. A background $\CS_g$ level-1 from the background integer quantum Hall state has been included. Adding a positive/negative mass gaps out the Majorana fermion and adds an extra gravitational Chern-Simons response of level $\pm 1/4$.
\begin{equation*}
    \cL_\text{Majorana}^>=\frac{3}{2}\CS_g,\qquad \cL_\text{Majorana}^<=\CS_g.
\end{equation*}
They recover the chiral central charge of the fIQH and MQH phases. This transition has another conjectural dual description of $\SO(3)_1$ coupled to a single free scalar~\cite{Metlitski:2016dht,Aharony2016Duality,Cordova2017Duality}. The condensation of the scalar Higgses the gauge theory to $\SO(2)_1$ describing the fIQH phase, and gapping the scalar results in $\SO(3)_1$ corresponding to the MQH phase.

The transition from the MQH to the bPf phase falls into a gauged Ising (Ising$^\ast$) universality, \textit{i.~e.}~Ising type transition coupled to a $\BZ_2$ gauge field, or the confinement transition of a $\BZ_2$ gauge field. One way to understand it is to recognise that bPf can be described by MQH coupled to a $\BZ_2$ gauge field, so its transition to MQH can be realised by Higgsing the $\BZ_2$ gauge field. We can also use the language of CS theory to understand the emergence of gauged Ising transition. The MQH and the bPf phase can be described by $\SO(3)_1\cong\SU(2)_2/\BZ_2$ and $\SU(2)_2$ respectively, and the difference is only a $\BZ_2$ gauge field. So Higgsing the $\BZ_2$ centre of the $\SU(2)$ gauge group drives a transition from the bPf phase to the MQH phase. In Appendix \ref{app:u2cs}, we discuss the Lagrangian of this confinement transition. This transition can be equivalently considered in terms of anyon condensation. In the bPf phase, although the $\psi$ anyon is fermionic, combined with the local fermion $f$, $f^\dagger\psi$ is bosonic with statistical angle $\theta_{f^\dagger\psi}=0$. Across the transition, this composite particle $\langle f^\dagger\psi\rangle\neq 0$ is condensed. Due to the braiding of the anyons $\psi$ and $\sigma$, $\sigma$ will also be confined, resulting in the MQH with trivial topological order. Since this transition only changes the neutral sector, the chiral central charge $c_-=3/2$ remains the same across the transition. This MQH-bPf transition can also be described as the adjoint QCD of $\SU(2)_1$ Chern-Simons theory coupled to a fermion in the adjoint representation of the gauge group~\cite{Gomis2017Adjoint}. When the fermion is gapped, it contributes a Chern-Simons level $\pm 1$ depending on the sign of the mass. This drives a transition between $\SU(2)_2$ topological order and a trivial topological order $\SU(2)_0$, which only differs from the MQH phase by an invertible TQFT. This transition is conjectured to be dual to $\rO(1)_{-3}$ coupled to a real scalar through the level-rank duality, which is equivalent to the Ising$^\ast$ universality.

Numerically, we verify these critical theories by inspecting the energy spectrum at the critical point. A quantum system on a sphere described by a CFT enjoys the state-operator correspondence where the scaling dimensions of the operators are proportional to the excited energies
\begin{equation*}
    \Delta_\Phi=\frac{v}{R}(E_\Phi-E_0)
\end{equation*}
where $\Phi$ is an arbitrary scaling operator and $v$ is a model-dependent constant that can be determined by calibrating the stress tensor $T^{\mu\nu}$ to $\Delta=3$. The spectrum of a CFT can be classified into multiplets of primaries and their descendants at integer spacing.

\begin{figure}[htbp]
    \centering
    \includegraphics[width=0.49\linewidth]{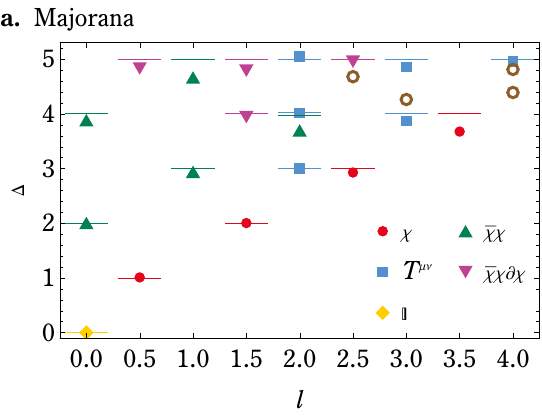}
    \includegraphics[width=0.49\linewidth]{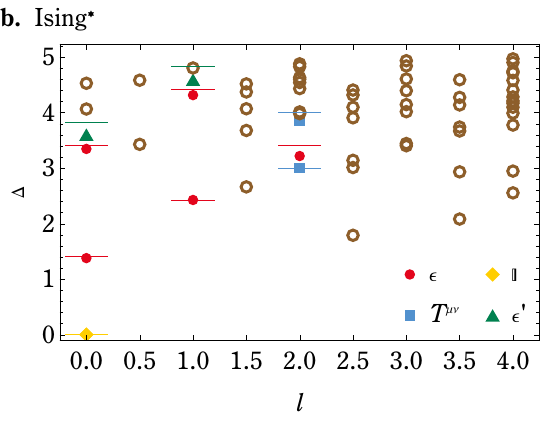}
    \caption{The rescaled energy spectrum at the conformal points (a) $t=0.3,\mu=0.0$ described by the free-Majorana-fermion CFT and (b) $t=0.1,\mu=2.25$ described by the Ising$^\ast$ CFT. We identify some multiplets of the lowest primaries in the spectra and the bars denote their expected value in the CFT. The spectra are calibrated by $\Delta_T=3$. The brown circles denote states not identified. In the calculation, we fix $N_{mf}=12$.}
    \label{fig:conf_spec}
\end{figure}

At the fIQH-MQH transition, we observe that most of the operators are close to integer levels and they can be classified into multiplets of the lowest primary operators in the free-Majorana-fermion CFT (Figure \ref{fig:conf_spec}a), \textit{viz.}~the fermion $\chi$ at $\Delta=1,l=1/2$, the mass field $\bar{\chi}\chi=\epsilon_{\alpha\beta}\chi^\alpha\chi^\beta$ at $\Delta=2,l=0$, the conserved stress tensor $T^{\mu\nu}=\tfrac{i}{2}\bar{\chi}_a\alpha(\gamma^{(\mu})^\alpha{}_\beta\partial^{\nu)}\chi^\beta$ at $\Delta=3,l=2$, $\bar{\chi}\chi\partial \chi$ at $\Delta=4,l=3/2$, \textit{etc}. We highlight that spectrum contains $l\in\BZ+\half$ states corresponding to fermionic operators. The free-Majorana-fermion CFT will be analysed in more detail in Section \ref{sec:majorana}.

At the MQH-bPf transition, although there are many non-conformal gapped excitations at relatively low energy at small system size, the low lying states for each integer $l\leq 2$ can be identified as operators in the Ising$^\ast$ CFT (Figure \ref{fig:conf_spec}b). The Ising$^\ast$ CFT consists all the $\BZ_2$-even operators in the Ising CFT. We can identify $\epsilon$ at $\Delta=1.412$ and its several descendants, the stress tensor $T^{\mu\nu}$ and $\epsilon'$ at $\Delta=3.830$ in the spectrum. The Ising$^\ast$ CFT will be discussed in more detail in Section \ref{sec:ising}.

\section{Free-Majorana-Fermion CFT}
\label{sec:majorana}

In this section, we focus on the free-Majorana-fermion CFT on the fuzzy sphere. In this CFT, the Majorana fermion $\chi$ has scaling dimension $\Delta=1$ and $\SU(2)$ spin $l=1/2$. The spectrum of scaling operator, including the primaries and descendants, can be determined with the help of harmonic oscillators~\cite{He2025Jun}. Here we list the scaling dimension $\Delta$ and spin $l$ of the low lying operators:
\begin{align*}
    \Delta&=1&l&=1/2\\
    \Delta&=2&l&=0,\ 3/2\\
    \Delta&=3&l&=1,\ 2,\ 5/2\\
    \Delta&=4&l&=0,\ 3/2,\ 2,\ 2,\ 3,\ 7/2\\
    &\hphantom{{}={}}\vdots
\end{align*}
where the replicated numbers denote degenerate operators. To choose an ideal parameter point, we optimise a cost function along the critical line defined as the root mean square of the deviations of the scaling dimension of all operators with $\Delta\leq\Delta_c=3$ from their expected values~\cite{Zhou2024Oct}. We find that in the region $0.2<t<0.4$, the cost function is relatively insensitive to the choice of $t$, and the critical $\mu_{c1}\approx 0$ up to a precision of $0.01$, which may be a hint for a possible enhanced symmetry in the absence of relative chemical potential. We therefore fix $t=0.3$ and $\mu_c=0$.

\begin{figure}[htbp]
    \centering
    \includegraphics[width=0.7\linewidth]{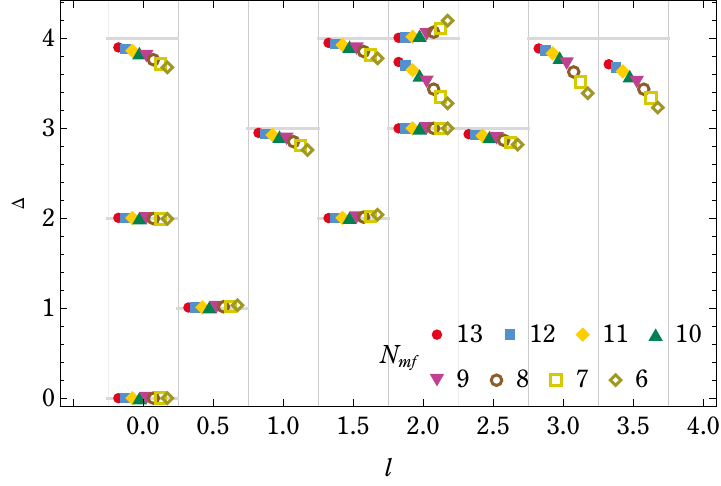}
    \caption{The spectrum with $\Delta\lesssim 4$ for various systems sizes in the free-Majorana-fermion CFT. The scaling dimensions are calibrated by the stress tensor with $\Delta_T=3$. The grey bars denote the theoretical value for comparison. In the calculation, we fix $t=0.3$ and $\mu=0$. The verticle gridlines with interval $0.5$ separates differnt spins, which are either integers or half-integers.}
    \label{fig:maj_spec}
\end{figure}

\begin{figure}[htbp]
    \centering
    \includegraphics[width=0.59\linewidth]{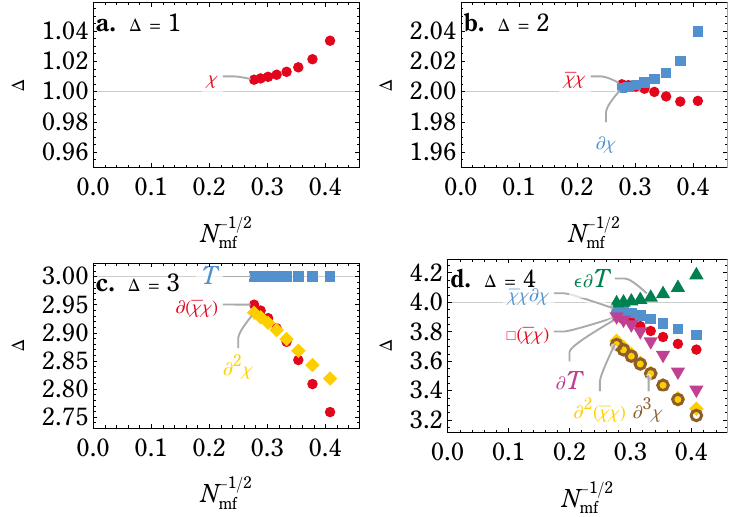}
    \includegraphics[width=0.39\linewidth]{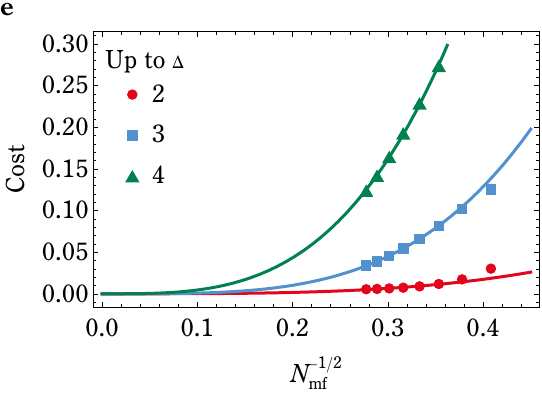}
    \caption{(a--d) The finite-size scaling of scaling dimensions of various operators in the free-Majorana-fermion CFT. The grey lines denote the theoretical value for comparison. (e) The finite-size scaling of cost function consisting of operators with $\Delta\leq\Delta_c=2,3,4$. In the calculation, we fix $t=0.3$ and $\mu=0$.}
    \label{fig:maj_scal}
\end{figure}

We find that the low lying spectrum agrees well with the theory (Figure \ref{fig:maj_spec}). All the states up to $\Delta=4$ are CFT states, and all the theoretically expected operators appear in the spectrum close to integer scaling dimensions. The discrepancy is within $2.5\%$ for all the $\Delta\leq 3$ operators and most $\Delta=4$ operators for our maximal accessible size $N_{mf}=13$. This precision is remarkable with almost no fine-tuning at such small size even compared with similar fuzzy-sphere model (\textit{e.~g.}~the free real scalar~\cite{He2025Jun}). With increasing system size, the scaling dimension gradually approaches to expected integer value (Figure \ref{fig:maj_scal}a--d). We quantitatively evaluate this through the cost function consisting of all the operators with $\Delta\leq\Delta_c=2,3,4$. We find these cost functions all scales to zero as $Q\sim N_{mf}^{-\alpha/2}$ where $\alpha=3.4,3.6,3.2$ respectively for $\Delta_c=2,3,4$ (Figure \ref{fig:maj_scal}e).

Having discussed the operator spectrum, we now move on to the correlation functions of the local operators, of which the most important is the Majorana fermion $\chi$ itself. We sketch the process here, and give the details in Appendix \ref{app:corr}. On the fuzzy sphere, the simplest gapless local fermionic operators are boson-fermion pairs $f^\dagger(\br)b(\br)$ and $b^\dagger(\br)f(\br)$. As $\chi$ is a two-component spinor, $f^\dagger b$ and $b^\dagger f$ respectively realises its two polarisations
\begin{align}
    \eta(\br)&=f^\dagger(\br)b(\br)=\alpha\chi^{(1/2)}(\br)+\dots\nonumber\\
    \eta^\dagger(\br)&=b^\dagger(\br)f(\br)=\bar{\alpha}\chi^{(-1/2)}(\br)+\dots,
\end{align}
where the two polarisations $\chi^{(s)}(\br)$ ($s=\pm 1/2$) of the spinor carry angular momentum $s$ under the $\SO(2)$ rotation along an axis that passes $\br$ (\textit{i.~e.}~spin-weight), and $\alpha$ is a coefficient that can be determined from the microscopic model. The conformal kinematics requires the two-point correlators to have the form
\begin{equation}
    \langle\chi^\alpha(x_1)\bar{\chi}_\beta(x_2)\rangle=\frac{(\gamma_\mu)^\alpha{}_\beta x_{12}^{\mu}}{(x_{12}^2)^{\Delta_\chi+1/2}},
\end{equation}
where $\Delta_\chi=1$, $x_{1,2}\in\BR^3$, $x_{12}=x_1-x_2$, $\alpha,\beta=\pm$ are spinor indices, and $\gamma$ are the Pauli matrices. Taking the polarisations $\chi^{(s)}(\br)$ on the sphere, their two-point function
\begin{equation}
    \langle 0|\chi^{\dagger\,(s_1)}(\br_1)\chi^{(s_2)}(\br_2)|0\rangle=\frac{2(-1)^{s_1}\delta_{s_1s_2}}{R^{2\Delta_\chi}(2\sin\frac{1}{2}\theta_{12})^{2\Delta_\chi}}
\end{equation}
has the same form as scalars on the sphere, where $\theta_{12}$ is the angular distance between the two points.

\begin{figure}[htbp]
    \centering
    \includegraphics[width=0.49\linewidth]{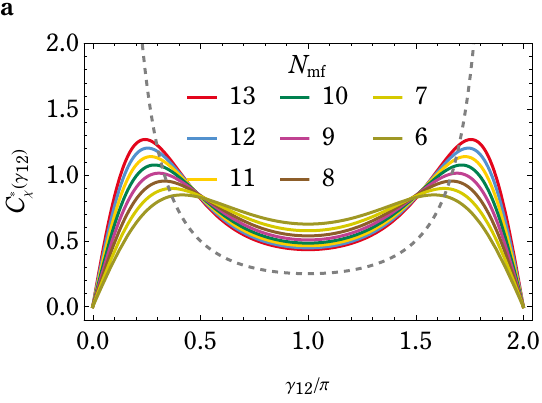}
    \includegraphics[width=0.49\linewidth]{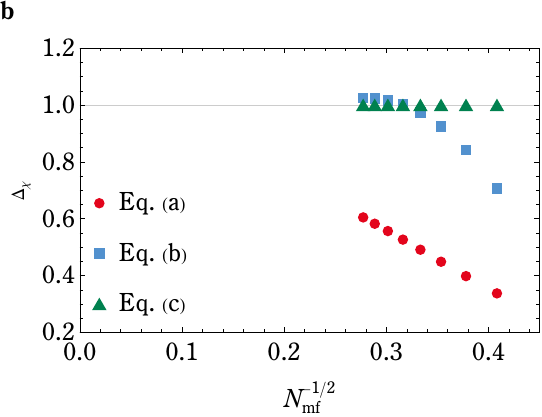}
    \caption{(a) The dimensionless correlation function of the Majorana fermion $\chi$ Eq.~\eqref{eq:corr_dl} at diffrent system sizes. The grey dashed line denotes $(2\sin\theta_{12}/2)^{-2}$ required by conformal symmetry. (b) The scaling dimension $\Delta_\chi$ extractd from the corrlation function through different ways in Eqs.~\eqref{eq:corr_dim} as a function of system size. In the calculation, we fix $t=0.3$ and $\mu=0$.}
    \label{fig:maj_corr}
\end{figure}

We calculate the dimensionless correlation function
\begin{equation}
    C_\chi^\ast(\theta_{12})=\tfrac{1}{2i}R^{2\Delta_\chi}\langle 0|\chi^{\dagger\,(1/2)}(\br_1)\chi^{(1/2)}(\br_2)|0\rangle.
    \label{eq:corr_dl}
\end{equation}
whose exact expression is given in Eq.~\eqref{eq:corr_dl_exact}, and plot the results in Figure \ref{fig:maj_corr}a. As the system size increases, the numerical result gradually approaches its theoretical value $(2\sin\theta_{12}/2)^{-2\Delta_\chi}$. From this correlation function, the scaling dimension $\Delta_\chi$ can be extracted in several ways
\begin{subequations}
\label{eq:corr_dim}
\begin{align}
    \left.C_\chi^\ast(\theta_{12})\right|_{\theta_{12}=\pi}&=2^{-2\Delta_\chi}\\
    \left.\frac{(C^\ast_\chi)''(\theta_{12})}{C^\ast_\chi(\theta_{12})}\right|_{\theta_{12}=\pi}&=\frac{\Delta_\chi}{2}\\
    \int_{-1}^1\rd\cos\theta_{12}\,\sin\frac{\theta_{12}}{2}C^\ast_\chi(\theta_{12})&=\frac{2^{2(1-\Delta_\chi)}}{3-2\Delta_\chi}.
    \label{eq:corr_dim_int}
\end{align}
\end{subequations}
We find the results all approach $\Delta_\chi=1$ with increasing system size (Figure \ref{fig:maj_corr}b). The results from the integral Eq.~\eqref{eq:corr_dim_int} has an especially small deviation within $10^{-3}$ at all system sizes. 

\section{Gauged Ising CFT}
\label{sec:ising}

In this section, we focus on the gauged Ising (\textit{i.~e.}~Ising$^\ast$) CFT on the fuzzy sphere. Gauging the $\BZ_2$ symmetry brings no change to the dynamics. The operator content of the gauged Ising CFT is the $\BZ_2$-even sector in the original Ising CFT, which includes $\epsilon$, $T^{\mu\nu}$, \textit{etc}. The $\BZ_2$-odd operators (\textit{e.~g.}~$\sigma$) of the original Ising CFT are no longer local operators in the gauged Ising CFT because they are not gauge invariant.

\begin{figure}[htbp]
    \centering
    \includegraphics[width=0.49\linewidth]{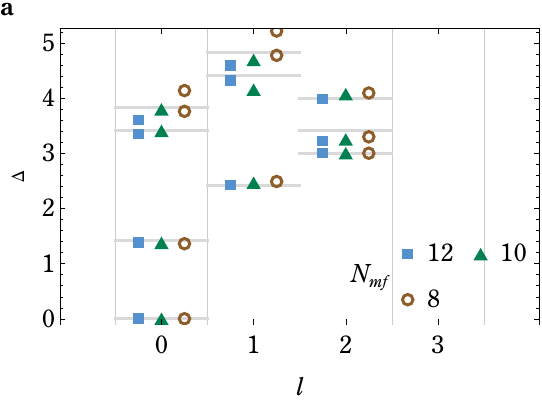}
    \includegraphics[width=0.49\linewidth]{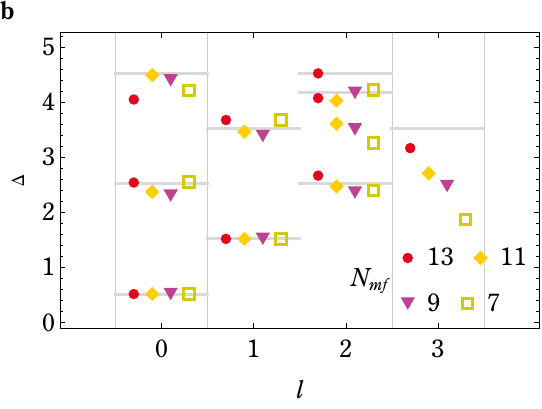}
    \caption{The spectrum of the identified CFT states with $\Delta<5$ for various systems sizes. The grey bars denote the theoretical value for comparison. (a) The bulk spectrum at even $N_{mf}$ corresponding to the $\BZ_2$-even part of the Ising CFT spectrum. The scaling dimensions are calibrated by the stress tensor with $\Delta_T=3$. (b) The Wilson-line end-point spectrum at odd $N_{mf}$ corresponding to the $\BZ_2$-odd part of the Ising CFT spectrum. The scaling dimensions are calibrated by setting $\Delta_\sigma=0.518$ and $\Delta_{\partial^\mu\sigma}=1.518$. In the calculation, we fix $t=0.1$ and choose $\mu$ as in Table~\ref{tbl:para_ising}.}
    \label{fig:ising_spec}
\end{figure}

Numerically, we can indeed identify $\epsilon,\epsilon',T^{\mu\nu}$ and their descendants up to $\Delta<5$ and $l\leq 2$ in the spectrum (Figure \ref{fig:ising_spec}a). In the calculation, we fix $t=0.1$ with a relatively optimal conformal symmetry, and choose $\mu$ for each system size by optimising a cost function consisting of $\epsilon,\partial^\mu\epsilon,\partial^\mu\partial^\nu\epsilon,\Box\epsilon$ and $T^{\mu\nu}$. The optimal $\mu$ for each size is listed in Table~\ref{tbl:para_ising} (the meaning of the right panel will be explained shortly). We note that although there are non-conformal states due to the lowest gapped excitations in the proximate bosonic Pfaffian phase, the low energy spectrum with $l\leq 2$ show relatively clear conformal signature. At large enough system size, the low energy specctrum should be dominated by CFT states and these gapped excitations will move higher, but at the sizes available, these excitations appear at relatively low energy at large $l$. As the bosonic Pfaffian state is only compatible with even $N_{mf}$, and the optimal conformal point drifts with system size, we leave the finite-size scaling analysis to future work.

\begin{table}[htbp]
    \centering
    \begin{tabular}{c|ccc|cccc}
        \hline\hline
        $N_{mf}$&$12$&$10$&$8$&$13$&$11$&$9$&$7$\\
        \hline
        $\mu_{c2}$&$2.254$&$2.230$&$2.356$&$2.290$&$2.318$&$2.373$&$2.396$\\
        \hline\hline
    \end{tabular}
    \caption{The optimal conformal point $\mu_{c2}$ for the Ising$^\ast$ CFT for each system size. In the calculation, we fix $t=0.1$.}
    \label{tbl:para_ising}
\end{table}

On the other hand, the $\BZ_2$-odd operators can be made gauge invariant by attaching a Wilson line (of $\BZ_2$ gauge field) to it. In other words, the $\BZ_2$-odd operators become end-point operators of a line defect in the gauged Ising CFT. Although the defect end-point operators arise generically in conformal line defects and have been discussed in various contexts~\cite{Cardy1989Defect,Oshikawa1996Defect,Zhou2024Jan,Lanzetta2025Bootstrap}, in the setting of $\BZ_2$ (or any discrete) gauge theory, the Wilson line defect is not only conformal but also topological. In the bPf-MQH transition, we can straightforwardly realise the topological Wilson line on the fuzzy sphere by leveraging the even-odd effect of $N_{mf}$.

We first note that the bosonic Pfaffian state is sensitive to the parity of particle number, and is only compatible with even $N_{mf}$. The choice of the odd $N_{mf}$, on the other hand, is equivalent to the insertion of a gauge charge at the centre of the sphere as the end-point of a Wilson line defect. The reason is that although the $\rU(1)_e$ symmetry of electric charge conservation decouples from the dynamics at the critical point, it still has topological contribution. First, odd $N_{mf}$ can be understood as the insertion of a $2\pi$-flux of $\rU(1)_e$ on top of even $N_{mf}$, \textit{i.~e.}, $q$ is increased by $1/2$. Second, the $2\pi$-flux insertion is bound to a $\BZ_2$ gauge charge. The binding of flux and gauge charge can be seen most evidently from the mutual Chern-Simons $(1/2\pi)\,a\,\rd A$ coupling of the the electromagnetic field $A$ and the dynamical gauge field $a$ (\textit{cf.}~Appendix \ref{app:u2cs}). Here $a$ is a $\rU(1)$ dynamical gauge field which is Higgsed to $\BZ_2$.\footnote{An equivalent way to see this Higgs mechanism is to impose $\rd a=0$ and $\oint_\mathcal{C}a=0$ or $\pi$, where $\mathcal{C}$ is a closed loop.} Its equation of motion asserts
\begin{equation*}
    Q_a=\frac{1}{2\pi}\Phi_A,
\end{equation*}
\textit{i.~e.}, upon the insertion of $2\pi$-flux of $A$, a $\BZ_2$ gauge charge is also inserted together with an electric charge. Due to the topological nature, the attached Wilson line preserves the $\SO(3)$ sphere rotation symmetry, which is different from the usual case of a non-topological conformal line. The defect end-point operators are the $\BZ_2$-odd part of the Ising CFT, and they can be observed from the eigenstates through the state-operator correspondence on the fuzzy sphere, as we will discuss below.

\begin{figure}[htbp]
    \centering
    \includegraphics[width=0.49\linewidth]{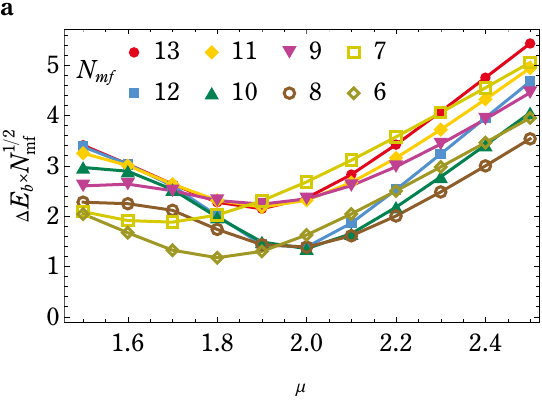}
    \includegraphics[width=0.49\linewidth]{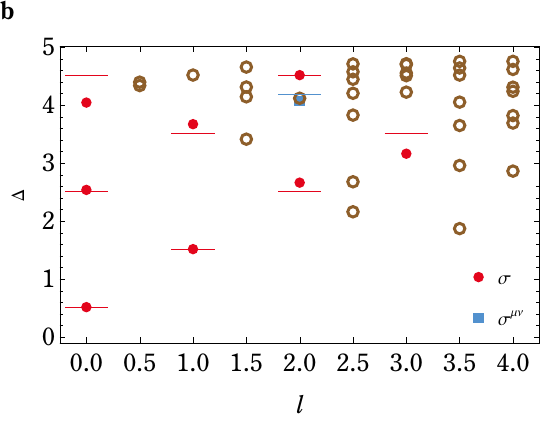}
    \caption{(a) The rescaled bosonic energy gap $\Delta E_bN_{mf}^{1/2}$ as a function of $\mu$ in vicinity of the Ising$^\ast$ transition for both even and odd $N_{mf}$ calculated at $t=0.1$. The lines with even and odd $N_{mf}$ intersect respectively at $\mu_{c2}\approx2.0$ except for the smallest system sizes $N_{mf}=6,7$. (b) The rescaled energy spectrum at conformal point $t=0.1,\mu=2.29$ calculated at $N_{mf}=13$ described by the Wilson-line end-point operators in the Ising$^\ast$ CFT. We identify some multiplets of the lowest primaries in the spectra and the bars denote their expected value in the CFT. The energy spectrum is rescaled such that $\Delta_\sigma=0.518$ and $\Delta_{\partial^\mu\sigma}=1.518$.}
    \label{fig:ising_def}
\end{figure}

We first confirm check the closure of the bosonic gap for the odd $N_{mf}$. The bosonic gap $\Delta E_b$ closes for the odd $N_{mf}$ at $\mu_{c2}\approx2.0$, similar to even $N_{mf}$, verified by overlapping $\Delta E_b\times N_{mf}^{1/2}$ (Figure \ref{fig:ising_def}a). Moreover, the value of $\Delta E_b\times N_{mf}^{1/2}$ for even $N_{mf}$ is approximately $0.7$ times that for odd $N_{mf}$. This is consistent with the expectation from the state-operator correspondence. In the gauge-neutral sector, the lowest two spin-$0$ operators are vaccuum $\BI$ and $\epsilon$ with a scaling dimension difference $1.412$; in the gauge-charged sector, they are $\sigma$ and $\Box\sigma$ with a scaling dimension difference $2$.

We then analyse the energy spectrum of the Wilson-line end-point. As the gauge charge is locked with $N_{mf}$, we cannot access the bulk spectrum and Wilson-line end-point spectrum at the same time and use the wave-function overlap scheme~\cite{Zou2021Defect,Zou2022Defect,Zhou2024Jan} to extract the scaling dimension of end-point operators. Instead, we fix the lowest end-point operator $\sigma$ at $\Delta=0.518$, and use the second lowest operator $\partial^\mu\sigma$ at $\Delta=1.518$ as the calibrator. The lowest operators up to $\Delta<5$ and $l\leq 3$ can be identified with $\BZ_2$-odd operators in the Ising CFTs, \textit{viz.}~$\sigma$ with $\Delta=0.518$ and its descendants, and $\sigma^{\mu\nu}$ with $\Delta=4.180$ (Figure \ref{fig:ising_def}b). Like for the gauge-neutral sector, we fix $t=0.1$ and optimise a cost function consisting $\sigma$, $\partial^\mu\sigma$, $\partial^\mu\partial^\nu\sigma$, $\Box\sigma$, $\Box\partial^\mu\sigma$ and $\sigma^{\mu\nu}$ for each system size to find the optimal $\mu$ (Table~\ref{tbl:para_ising}, right panel). The spectra of identified conformal operators for each system size are plotted in Figure \ref{fig:ising_spec}b. They agree considerably well with the odd sector of Ising CFT.

\section{Correspondence with QFT Lagrangians}
\label{sec:lag}

In Ref.~\cite{He2025Jun}, an analogy between the fuzzy-sphere models and the field theories have been noted for the free and critical scalars. 
In this section, we further extend this correspondence to free and interacting Majorana fermions. 

\begin{figure}
    \centering
    \includegraphics[width=0.24\linewidth,valign=t]{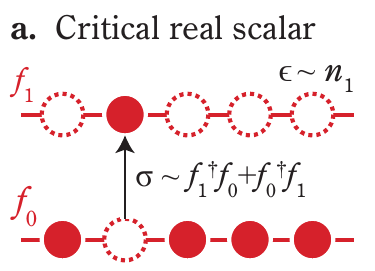}
    \includegraphics[width=0.24\linewidth,valign=t]{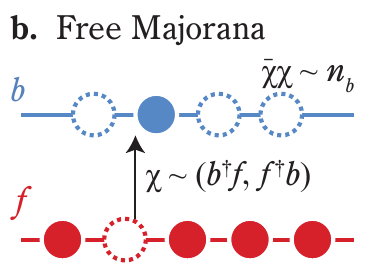}
    \includegraphics[width=0.40\linewidth,valign=t]{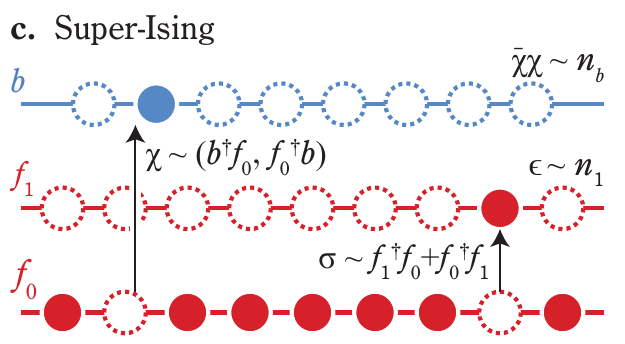}
    \caption{The illustration of the correspondence between fuzzy-sphere set-up and the field-theory operators in (a) the critical real scalar (Ising), (b) the free Majorana fermion, and (c) the super-Ising theory.}
    \label{fig:demo_lag}
\end{figure}

For a critical real scalar (Ising), the set-up contains two fermionic flavours $f_{0,1}$ that are respectively even and odd under the $\BZ_2$ symmetry. The $f_0$ flavour can be identified as a `vacuum' or reference flavour --- the CFT ground state can be regarded as a perturbation from the reference state where all the $f_0$ particles are occupied. As an evidence, the CFT ground state has a substantial overlap with the reference state. As the reference state is a direct product state in the orbital space, the orbital-space entanglement entropy is very small for the CFT ground state. The orbital-space entanglement entropy is measured through a bi-partition between the $m<0$ orbitals and the $m\geq 0$ orbitals. Correspondingly, the $f_1$ flavour is identified as an `excited state.' The hopping between the two flavours is identified as the elementary field $\sigma$, and the density of `excitations,' or particles in the $f_1$ flavour, is identified as the mass field $\epsilon=\sigma^2$ up to a constant
\begin{align}
    \sigma&\sim n_x=f_1^\dagger f_0+f_0^\dagger f_1,&\epsilon+\text{const.}&\sim n_1=f_1^\dagger f_1.
\end{align}
Here the `$\sim$' denotes a phenomelogical correspondence between a field theory operator and a fuzzy-sphere operator, or between a term in the field-theory Lagrangian and a term in the fuzzy-sphere Hamiltonian.\footnote{The correspondence holds more precisely between the field-theory Hamiltonian and fuzzy-sphere Hamiltonian. Since the Lagrangian and Hamiltonian can be easily converted, we pick the Lagrangian as a more familiar representative.} The correspondence is illustrated in Figure~\ref{fig:demo_lag}a. All the terms in the Hamiltonian 
\begin{equation}
    H=\int\rd^2\br\left[ n_e(\br)^2+Un_x(\br)\nabla^2n_x(\br)-\mu_1n_1(\br)\right],
\end{equation}
besides a electric density interaction of $n_e=f_0^\dagger f_0+f_1^\dagger f_1$ that suppresses the charge fluctuation, has a direct intepretation to the field theory 
\begin{align}
    \cL&=\frac{1}{2}(\partial^\mu\sigma)^2-\frac{1}{2}m^2\sigma^2-\frac{\lambda}{4!}\sigma^4.
\end{align}
Similar to the lattice models like the transverse-field Ising model, the Ising-type density-density interaction can be intepreted as the $\phi^4$-theory at a certain mass $m_0$, and the chemical potential fine tunes the mass term
\begin{align}
    Un_x\nabla^2n_x&\sim\frac{1}{2}(\partial^\mu\sigma)^2-\frac{1}{2}m_0^2\sigma^2-\frac{\lambda}{4!}\sigma^4,&\mu_1n_1&\sim\frac{1}{2}(m^2-m_0^2)\sigma^2.
\end{align}

\begin{figure}[htbp]
    \centering
    \includegraphics[width=0.33\linewidth]{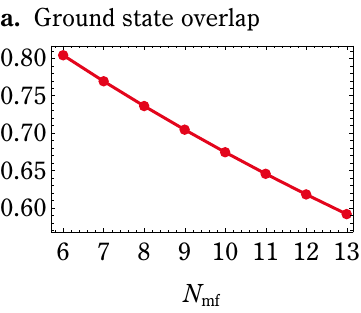}
    \includegraphics[width=0.33\linewidth]{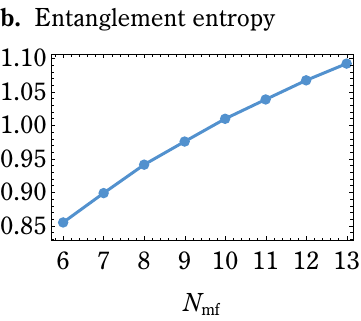}
    \caption{Properties of the CFT ground state of the free Majorana fermion: (a) The overlap $|\langle\text{ref}|0_\text{CFT}\rangle|$  between the CFT ground state and the reference state \eqref{eq:st_ref} where all fermions are occupied; (b) the orbital-space entanglement entropy. The data are measured at $t=0.3,\mu_{c1}=0$ and plotted as a function of the system size $N_{mf}$.}
    \label{fig:maj_pheno}
\end{figure}

While the correspondence between field-theory Lagrangian and the fuzzy-sphere model is to some extent vague for the critical scalar, it can be seen much more clearly for the free Majorana fermion. The set-up now contains one fermionic `reference' flavour $f$ and a bosonic `excitation' flavour $b$. The CFT ground state has a substantial overlap $|\langle\text{ref}|0_\text{CFT}\rangle|$ with the reference state where all fermions are occupied (Figure~\ref{fig:maj_pheno}a)
\begin{equation}
    |\text{ref}\rangle=\prod_{m=-q}^qf_m^\dagger|\emptyset\rangle
    \label{eq:st_ref}
\end{equation}
and its orbital-space entanglement entropy is small (Figure~\ref{fig:maj_pheno}b). The conversion between the two flavours are identified as the elementary spinor field $\chi$, and the density of `excitations,' or particles in the $b$ flavour, is identified as the mass field $\bar{\chi}\chi$ up to a constant (Figure~\ref{fig:demo_lag}b)
\begin{align}
    \chi&\sim(\eta,\eta^\dagger)^\rT=(f^\dagger b,b^\dagger f)^\rT,&\bar{\chi}\chi+\mathrm{const.}&\sim n_b=b^\dagger b
\end{align}
The terms in the Hamiltonian can again be intepreted as the kinetic term and the mass term
\begin{align}
    \frac{1}{2}\bar{\chi}\,i\slashed{\partial}\chi&\sim t\left(\eta D_+\eta+\hc\right),&\frac{1}{2}m\bar{\chi}\chi\sim\mu n_b
\end{align}
where $D_+=D_\theta+iD_\phi$ in addition to an electric density interaction. 

Notably, the pair conversion term $H_t$ bears exactly the same form as the kinetic term in the Hamiltonian of free Majorana fermion upon radial quantisation.
\begin{equation}
    H_t\propto \int\rd^2\br\,i\epsilon^{\mu\nu\rho}\hat{n}_\mu\bar{\chi}_\alpha(\br)(\gamma_\nu)^\alpha{}_\beta\partial_\rho\chi^\beta(\br)=\int\rd^2\br\,i\bar{\chi}(\br)(\hat{\mathbf{n}}\times\gamma\cdot\partial)\chi(\br).
    \label{eq:hmt_maj_qft}
\end{equation}
As an evidence for this correspondence, the critical point locates at $\mu_c=0$ to a high precision where the mass term vanishes.

We shall show that these ingredients suffice to construct the interacting theories. The simplest one is a Majorana fermion $\chi$ coupled to a real scalar $\sigma$ through a Yukawa interaction~\cite{Grover2013SuperIsing,Gross1974}
\begin{equation}
    \cL[\chi,\sigma]=\frac{1}{2}\bar{\chi}\,i\slashed{\partial}\chi+\frac{1}{2}(\partial^\mu\sigma)^2-\frac{1}{2}m_\sigma^2\sigma^2-\frac{\lambda}{4!}\sigma^4-\frac{g}{2}\sigma\bar{\chi}\chi.
    \label{eq:lag_suis}
\end{equation}
The Yukawa interaction locks the Ising $\BZ_2$ symmetry to the space-time parity. As a result, this theory has no global symmetry, and $\sigma$ is odd under space-time parity. When the parity is not manifest, two additional perturbations should also be considered, namely the fermion mass and the polarisation term
\begin{equation*}
    m_\chi\bar{\chi}\chi+h\sigma.
\end{equation*}
In the infra-red, the Lagrangian flows to a $\cN=1$ super-conformal field theory (SCFT) known as the super-Ising theory. We will introduce its property in more detail in Section~\ref{sec:superising}.

The corresponding set-up on the fuzzy sphere contains two fermionic flavours $f_0,f_1$ and a bosonic flavour $b$. We set the bosons and fermions to have a mismatch $q_f-q_b=1/2$ in the angular momenta. We set the total filling to be $1$, \textit{i.~e.}~$N_e=N_{mf}$. Similar to the critical real scalar and the free Majorana fermion, we interpret the $f_0$ flavour as the reference flavour, $f_1$ as the scalar excitation and $b$ as the fermion excitation. Hence, the elementary fields $\chi$ and $\sigma$ are realised as the hoppings from the reference fermion flavour $f_0$ to respectively the boson flavour $b$ and the other fermion flavour $f_1$ (Figure~\ref{fig:demo_lag}c)
\begin{subequations}
\label{eq:op_suis}
\begin{align}
    \eta(\br)&=b^\dagger(\br)f_0(\br),&\chi(\br)&\sim(\eta(\br),\eta^\dagger(\br))^\rT\\
    n_x(\br)&=f_1^\dagger(\br)f_0(\br)+f_0^\dagger(\br)f_1(\br),&\sigma(\br)&\sim n_x(\br)
\end{align}
The fermion and scalar bilinears $\bar{\chi}\chi$ and $\sigma^2$ are realised using the density operators 
\begin{align}
    n_b&=b^\dagger b\sim\bar{\chi}\chi+\text{const.},&n_1=f_1^\dagger f_1\sim\sigma^2+\text{const.}
\end{align}
\end{subequations}
up to a constant. Using these ingredients, we can translate the terms in the Lagrangian to four-particle interactions in the fuzzy-sphere model. We start with the terms that describe the critical scalar and the free Majorana fermion
\begin{subequations}
\begin{align}
    \frac{1}{2}\bar{\chi}\,i\slashed{\partial}\chi&\sim t\left( \eta(\br)D_+\eta(\br)+\hc \right)\\
    \frac{1}{2}(\partial^\mu\sigma)^2-\frac{1}{2}m_0^2\sigma^2-\frac{\lambda}{4!}\sigma^4&\sim Un_x(\br)\nabla^2n_x(\br).
\end{align}
 The Yukawa coupling can be expressed as a contact interaction between $n_b$ and $n_x$
\begin{equation}
    \frac{g}{2}\sigma\bar{\chi}\chi\sim gn_x(\br) n_b(\br).
\end{equation}
The terms left are the scalar mass, and the two additional terms due to the absence of parity symmetry
\begin{align}
    \frac{1}{2}(m_\sigma^2-m_0^2)\sigma^2&\sim \mu_1 n_1,&\frac{1}{2}m_\chi\bar{\chi}\chi&\sim\mu_b n_b,&h\sigma&\sim hn_x.
\end{align}
\end{subequations}
Combined with an electric density interaction $n_e^2$ where $n_e=f_0^\dagger f_0+f_1^\dagger f_1+b^\dagger b$, the total Hamiltonian reads
\begin{multline}
    H=\int\rd^2\br\,\Big[n_e^2(\br)+t\left(\eta(\br)D_+\eta(\br)+\hc\right)+Un_x(\br)\nabla^2n_x(\br)+gn_x(\br) n_b(\br)\\
    -h n_x(\br)-\mu_1 n_1(\br)-\mu_bn_b(\br)\Big].
    \label{eq:hmt_suis}
\end{multline}
It contains six parameters $t,U,g,h,\mu_1$ and $\mu_b$. The latter three $h,\mu_1$ and $\mu_b$ controls the three relevant perturbations $\sigma,\epsilon=\sigma^2$ and $\sigma'=\bar{\chi}\chi$, so that we can fix the values of $t,U,g$ and fine tune $h,\mu_1,\mu_b$ to reach the super-Ising fixed point.\footnote{In Eq.~\eqref{eq:hmt}, we write the Hamiltonian in terms of the pseudo-potentials, while it is written in terms of local operators in Eq.~\eqref{eq:hmt_suis} for convenience. The difference produces a constant conversion factor.}

\section{Super-Ising SCFT}
\label{sec:superising}

We have constructed the model to realise the super-Ising theory on the fuzzy sphere. In this section, we present the numerical evidence for the emergence of super-conformal symmetry and the agreement with conformal bootstrap.

We first discuss the supersymmetry in the field-theory Lagrangian~\eqref{eq:lag_suis}. In the infra-red, it flows to a super-conformal field theory (SCFT) with super-charge $\cN=1$ known as the super-Ising theory. The supersymmetry is manifest in the Lagrangian at $\lambda=3g^2$ and $m_\sigma=0$. The action can be organised in terms of a super-field $\Sigma(x,\theta)=\sigma+\bar{\theta}\chi+\frac{1}{2}\bar{\theta}\theta\epsilon$ where $\theta$ and $\bar{\theta}$ are Grassmannian co-ordinates. The $\epsilon$ is an auxilliary field coupled to $\sigma^2$. By the equation of motion $\epsilon=\sigma^2$. The Poincar\'e super-charge $Q^\alpha$ acts on the fields as 
\begin{align*}
    [Q^\alpha,\sigma]&=\chi^\alpha,&
    \{Q^\alpha,\chi^\beta\}&=-(\gamma^\mu)^{\alpha\beta}\partial_\mu\sigma-\epsilon^{\alpha\beta}\epsilon,&
    [Q^\alpha,\epsilon]&=(\gamma^\mu)^{\alpha\beta}\partial_\mu\chi_\beta.
\end{align*}

The super-conformal group is homomorphic to a super-Lie group $\OSp(1|4)$~\cite{Gates2001,Park2000}. More precisely, the (Lorentzian) conformal transformations acting on fermions give the Lie group $\mathrm{Spin}(3,2) \cong \Sp(4,\BR)$ of real symplectic transformations. Besides that, the super-Lie algebra also contains two Grassmannian generators, \textit{i.~e.}~the Poincar\'e super-charge $Q^\alpha$ and the conformal super-charge $S^\alpha$, which extend the conformal group to $\OSp(1|4,\BR)$. These generators organise several conformal multiplets into a super-conformal multiplet in a similar way as how the conformal symmetry orgainses the scaling operators into conformal multiplets~\cite{Cordova2019}. Similar to $P^\mu$ and $K^\mu$ that act as ladder operators of $D$ increasing and decreasing the scaling dimension $\Delta$ by $1$, $Q^\alpha$ and $S^\alpha$ act as `half-ladder' operators changing $\Delta$ by $\pm1/2$. The conformal primaries are classified into super-conformal primairies that are the lowest-weight state of $Q^\alpha$, and the super-conformal descendants by acting $Q^\alpha$ on the super-primaries. Due to the odd Grassmannian parity of $Q^\alpha$, each super-multiplet contains at most 4 conformal multiplets. As an example, the Majorana fermion $\chi^\alpha=Q^\alpha\sigma$ and $\epsilon=\bar{Q}_\alpha Q^\alpha\sigma$ are now super-conformal descendants of the super-conformal primary $\sigma$, and their scaling dimensions are not independent $\Delta_\chi=\Delta_\sigma+1/2$, $\Delta_\epsilon=\Delta_\sigma+1$. The operator $\bar{Q}Q$ is odd under space-time parity, so $\sigma$ and $\epsilon$ has opposite space-time parity. 

The super-Ising SCFT has been studied extensively by perturbative calculation~\cite{Fei:2016sgs,Mihaila:2017ble} and conformal bootstrap~\cite{Rong2018SuperIsing,Atanasov2018SuperIsing,Atanasov2022SuperIsing}. The two relevant super-primaries $\sigma,\sigma'=\bar{\chi}\chi$ has been determined to a high precision. They are parity-odd Lorentz scalars and have scaling dimensions $\Delta_\sigma=0.584,\Delta_{\sigma'}=2.887$~\cite{Atanasov2022SuperIsing}. Besides that, a spin-$2$ parity-even operator $T'^{\mu\nu}$ at $\Delta_{T'}\approx 3.28$ has been observed in the extremal spectrum~\cite{Atanasov2022SuperIsing}. 

\begin{figure}[htbp]
    \centering
    \includegraphics[width=0.5\linewidth]{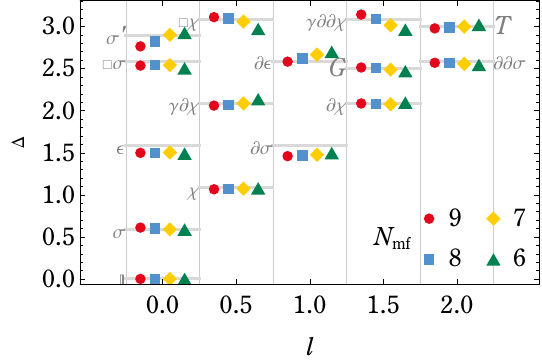}
    \caption{The low-lying spectrum of the super-Ising SCFT with $\Delta\lesssim 3$ and $l\leq 2$ for different system sizes. The scaling dimensions are calibrated by optimising the cost function. The grey bars denote the bootstrap values. The names of the operators are marked beside the bars. In the calculation, we fix $t=1.5,U=0.25,g=1$ and take $h,\mu_1,\mu_b$ as in Table~\ref{tbl:para_suis}.}
    \label{fig:suis_fss}
\end{figure}

To verify that our ansatz~\eqref{eq:hmt_suis} realises the super-Ising SCFT, we carry out an ED calculation up to $N_{mf}=9$. We fix $t=1.5$, $U=0.25$ and $g=1$ and determine $h$, $\mu_1$ and $\mu_b$ for each $N_{mf}$ by optimising a cost function defined as the root mean square of the relative deviations of the scaling dimension of all 14 operators with $\Delta<3.1$ and $l\leq 2$ (listed in Figure~\ref{fig:suis_fss}) from the values obtained from the conformal bootstrap~\cite{Atanasov2022SuperIsing}. The conformal points for $N_{mf}=9,8,7,6$ are listed in Table~\ref{tbl:para_suis}. The optimal parameters remain relatively stable with the system size. The low-lying spectrum for various $N_{mf}$ are plotted in Figure~\ref{fig:suis_fss}. For each $N_{mf}$, the lowest operators remain close to the expected values. The cost functions remain between $2\%$ and $3\%$. 

\begin{table}[htbp]
    \centering
    \begin{tabular}{c|ccc}
        \hline\hline
        $N_{mf}$ & $h$ & $\mu_1$ & $\mu_b$ \\ 
        \hline
        $9$ & $0.0680$ & $0.0809$ & $0.0776$ \\
        $8$ & $0.0690$ & $0.0792$ & $0.0757$ \\
        $7$ & $0.0698$ & $0.0774$ & $0.0722$ \\
        $6$ & $0.0662$ & $0.0793$ & $0.0615$ \\
        \hline\hline
    \end{tabular}
    \caption{The optimal conformal point of the super-Ising SCFT for different system sizes $N_{mf}$. In the calculation, we fix $t=1.5,U=0.25,g=1$.}
    \label{tbl:para_suis}
\end{table}

\begin{figure}[htbp]
    \centering
    \includegraphics[width=0.8\linewidth]{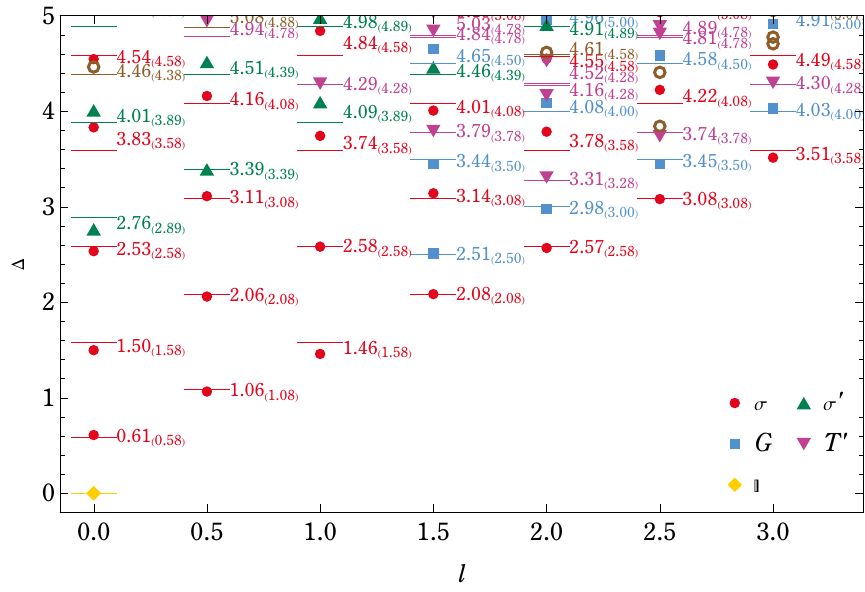}
    \caption{The spectrum of super-Ising field theory up to $\Delta\approx 5$ and $l=3$ at $N_{mf}=9$. The symbols marks the measured scaling dimensions and the bars marks the expected values from conformal bootstrap. Most states are identified in the lowest super-conformal multiplets. Each identified state is labelled by the measured and expected (in parentheses in subscript) scaling dimensions. The brown circles denote higher super-conformal states and three unidentified states. Higher spectrum is given in Appendix~\ref{app:suis_full}. In the calculation, we fix $t=1.5,U=0.25,g=1$ and take $h,\mu_1,\mu_b$ as in Table~\ref{tbl:para_suis}.}
    \label{fig:suis_spec}
\end{figure}

Going to slightly higher part of the spectrum (Figure~\ref{fig:suis_spec}), almost all states at $\Delta\lesssim 5,l\leq 3$ can be classified into super-multiplets of $\sigma$, $G$, $\sigma'$ and $T'$. The scaling dimensions of these operators are close to bootstrap results. We note that states with $\Delta>3.1$ are not included as inputs in the optimisation of the conformal point; nevertheless, they exhibit good consistency with the spectrum of the super-Ising theory. In Appendix~\ref{app:suis_full}, we give a full spectrum up to $\Delta=6.5$ and identify four more super-multiplets $\epsilon'',T_-,[\sigma\chi]_{l=7/2}$ and $[\sigma\sigma]_{l=4}$ from the bootstrap results. We have also confirmed a primary $G'^{\mu\alpha}$ at $l=3/2$ and $\Delta\approx5.28$ that is subject to large numerical uncertainty in the conformal bootstrap~\cite{Atanasov2022SuperIsing}.

Two key pieces of evidence for the emergent super-conformal symmetry are the supersymmetry Noether current and the super-conformal multiplet structure. For $\cN=1$ SCFTs, apart from the stress tensor, there exists another conserved supersymmetry current $G^{\alpha\mu}$ whose spatial integral gives the super-charge $Q^\alpha$. It is fermionic, has Lorentz spin $l_G=3/2$, and its scaling dimension saturates the unitarity bound $\Delta_G=5/2$.

\begin{figure}[htbp]
    \centering
    \includegraphics[width=0.49\linewidth]{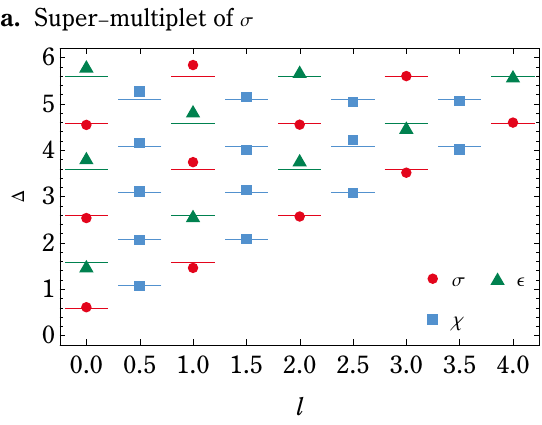}
    \includegraphics[width=0.49\linewidth]{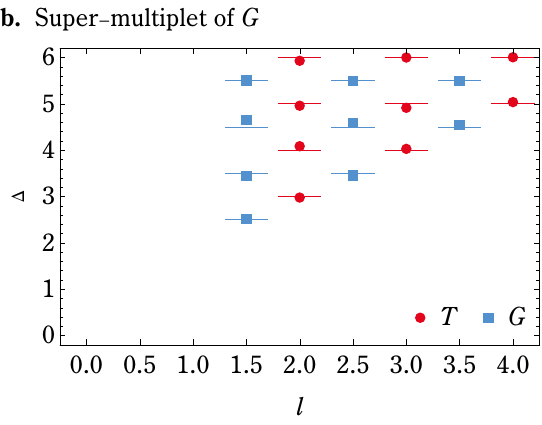}
    \includegraphics[width=0.49\linewidth]{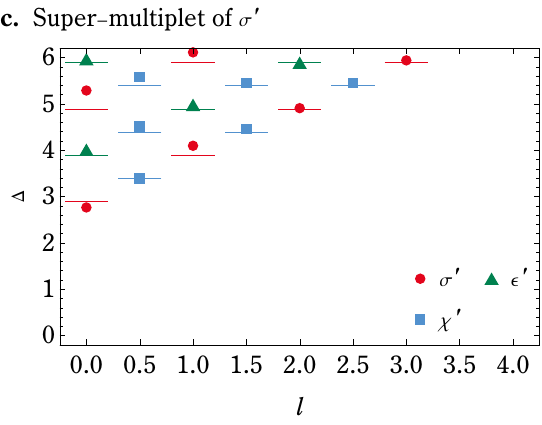}
    \includegraphics[width=0.49\linewidth]{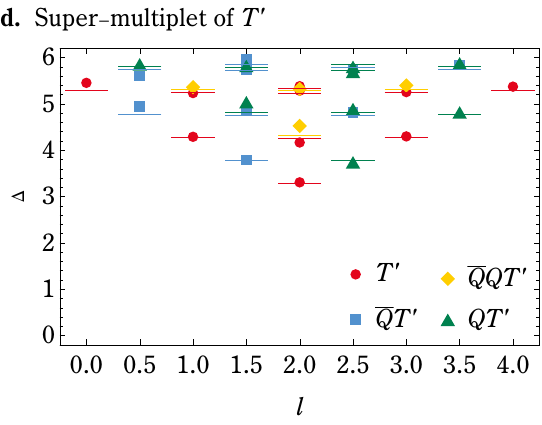}
    \caption{The super-conformal multiplets of the lowest super-conformal primaries (a) $\sigma$, (b) $G$, (c) $\sigma'$, and (d) $T'$ at $N_{mf}=9$ and up to $\Delta=6$ and $l=4$. The symbols marks the measured scaling dimensions and the bars marks the expected values from conformal bootstrap. The states in each super-conformal multiplets are organised into conformal multiplets. The degenerate states are strongly mixed and are assigned arbitrarily. In the calculation, we fix $t=1.5,U=0.25,g=1$ and take $h,\mu_1,\mu_b$ as in Table~\ref{tbl:para_suis}.}
    \label{fig:suis_mult}
\end{figure}

The super-conformal symmetry organises the scaling operators into super-conformal multiplets. Acting $Q^\alpha$ on a super-conformal primary gives its super-conformal descendants. The action alters the fermion parity, changes the Lorentz spin by $1/2$ and increases the scaling dimension by $1/2$. For each low-lying super-primary, we can find numerically all of its super-conformal descendants and their conformal descendants up to $\Delta\lesssim 6$ and $l\leq 4$ (Figure~\ref{fig:suis_mult}). Their scaling dimensions are at half-integer spacings as expected. In the following, we explain the structure of the super-multiplets:

For a general super-conformal primary $\Phi$ with Lorentz spin $0$, its super-multiplet contains three conformal primaries, \textit{viz.}~the bosonic $\Phi$, fermionic $Q^\alpha\Phi$ with spin-$1/2$ and $\Delta_\Phi+1/2$, and bosonic $\bar{Q}_\alpha Q^\alpha\Phi$ with spin-$0$ and $\Delta_\Phi+1$. As an example, the primaries $\sigma$, $\chi=Q\sigma$ and $\epsilon=\bar{Q}Q\sigma$ forms a super-conformal multiplet and their scaling dimensions are locked $\Delta_\chi=\Delta_\sigma+1/2,\Delta_\epsilon=\Delta_\sigma+1$ (Figure~\ref{fig:suis_mult}a). For a spin-$l$ primary $\Phi^{\alpha_1\dots\alpha_{2l}}$ (where each Lorentz index is expressed in terms of two spinor indices), \textit{e.~g.}~the spin-$2$ $T'_{\mu\nu}$, its super-multiplet contains $\Phi$, $Q\Phi=Q^{\alpha_0}\Phi^{\alpha_1\dots\alpha_{2l}}$, $\bar{Q}\Phi=\bar{Q}_{\alpha_1}\Phi^{\alpha_1\dots\alpha_{2l}}$ with spin-$(l\pm1/2)$ and $\Delta_\Phi+1/2$, and $\bar{Q}Q\Phi$ with spin-$l$ and $\Delta_\Phi+1$ (Figure~\ref{fig:suis_mult}d). States from different conformal multiplets can be degenerate protected by the super-conformal symmetry. \textit{E.~g.}, we have observed a nearly degeneracy of the conformal primary $\bar{Q}_\alpha Q^\alpha T'^{\mu\nu}$ and the conformal descendant $\epsilon^{\mu\rho\sigma}\partial_\rho T'^\nu{}_\sigma$ at $\Delta\approx 4.28$ and $l=2$, although subject to level repulsion~\cite{Laeuchli2025}. An exception to this structure is the shortening of the super-current super-multiplet due to the conservation $\bar{Q}_\alpha G^{\alpha\mu}=0$. It only contains the supersymmetry current $G^{\alpha\mu}$ and the stress tensor $T^{\mu\nu}=Q^\alpha(\gamma^\mu)_{\alpha\beta}G^{\nu\beta}$ as its super-descendant (Figure~\ref{fig:suis_mult}b).

\section{Discussion}
\label{sec:discussion}

In this paper, we have studied the free Majorana fermion, the gauged Ising CFT and the super-Ising SCFT on the fuzzy sphere. For the free Majorana fermion, we consider a set-up consisting of one bosonic and one fermionic lowest Landau level with a $1/2$ angular momentum mismatch, and allow the pair conversion between them. This microscopic realisation of the spin-statistics constraint expected in the IR is an interesting subject for future work. By tuning a relative chemical potential between a fermionic integer quantum Hall (fIQH) phase at high fermion density and a bosonic Pfaffian (bPf) phase ($\SU(2)_2$ Chern-Simons theory) at high boson density, we find an intermediate phase described by a Majorana quantum Hall (MQH) state with Kitaev index $\nu_K=3$. The three phases are separated by two continuous phase transition. Supported by numerical and theoretical evidence, we show that the fIQH-MQH transition is described by a free-Majorana-fermion CFT. Using state-operator correspondence, we find that the energy spectrum agrees well with the operator content of free Majorana fermion up to $\Delta\leq 4$. The local fermion operator is realised as the bilinear $f^\dagger(\br)b(\br)$ and $\hc$ on the sphere, whose two-point correlation function matches the expected conformal correlator. On the other hand, the MQH-bPf transition is described by a gauged Ising (Ising$^\ast$) CFT. Its spectrum agrees with the $\BZ_2$-even sector of the Ising CFT. The transition is only consistent with even $N_{mf}$. We find that odd $N_{mf}$ is equivalent to the insertion of a $\BZ_2$ gauge charge, and the spectrum corresponds to the Wilson-line end-point operators of the gauged Ising CFT, \textit{i.~e.}~the $\BZ_2$-odd sector of the Ising CFT. 

Furthermore, we have established a correspondence between the fuzzy-sphere model Hamiltonians and the field-theory Lagrangians. In this correspondence, a flavour serves as a reference state and the other serves as an `excitation' flavour. The elementary fields that constitute the field theory are identified as hopping between them. Using this framework, we have translated an interacting fermionic CFT --- the super-Ising theory, in which a Majorana fermion $\chi$ is coupled to a real scalar $\sigma$ through a Yukawa interaction --- into a fuzzy-sphere model with two fermionic flavours $f_{0,1}$ and one bosonic flavour $b$. The elementary fields are identified with the hopping operators $\sigma=f_0^\dagger f_1+f_1^\dagger f_0$ and $\chi=(b^\dagger f_0,f_0^\dagger b)^\rT$. The three tuning parameters admit direct physical intepretations as the relevant operators $\sigma$, $\epsilon=\sigma'$ and $\sigma'=\bar{\chi}\chi$. We have provided numerical evidence for the emergence of super-conformal symmetry and the agreement with the conformal bootstrap results. The super-conformal multiplet structure with half-integer spacing is clearly observed and most of the low-lying states are identified as SCFT operators.

Our work represents the first fuzzy-sphere realisation of CFTs with critical local fermions and constitutes a step towards studying renormalisable Lagrangians of quantum field theories on the fuzzy sphere. Most previous fuzzy-sphere works were limited to realising universality classes by constructing Hamiltonian with correct symmetry and anomaly and relying on RG flow to reach the desired conformal fixed point in the IR. The free Majorana fermion, the super-Ising theory, and the free and critical real-scalar models in Ref.~\cite{He2025Jun} offer a natural and versatile playground for studying the connection between fuzzy-sphere constructions and quantum field theories. In particular, our fuzzy-sphere Majorana-fermion model has several notable features which deserve further exploration:
\begin{enumerate}
    \item The critical point lies at $\mu_{c1}=0$ with high precision over a wide range of $t$, suggesting the presence of an enhanced symmetry that forbids the chemical-potential perturbation in the free-Majorana-fermion CFT.
    \item At the critical point $\mu_{c1}=0$, the Hamiltonian closely resembles the familiar free Majorana fermion QFT: Apart from the charge density interaction term $n_e^2$, it takes the same form as the field theory~\eqref{eq:hmt_maj_qft}.
    \item Analogous to the density operators, the angular components of the local fermion $\eta^{(1/2)}_{lm}=\int\rd^2\br\,\bar{Y}^{(1/2)}_{lm}(\br)\allowbreak f^\dagger(\br)b(\br)$ may exhibit an algebraic structure similar to that of the density operators. It would be particularly interesting to determine whether this algebra is a deformation of the fermionic harmonic oscillator, analogous to the free scalar on the fuzzy sphere~\cite{He2025Jun}. Following this line, the ground state could admit a description by a wave-function ansatz resembling a coherent state of $\eta$, as in the case of the free scalar.
\end{enumerate}
The small orbital-space entanglement entropy suggests that DMRG calculations could access larger system sizes and provide more definitive answers to these questions. Understanding these features may guide us to generalise the fuzzy-sphere approach to other interacting Lagrangians, \textit{e.~g.}~the gauge theories that require the engineering of spin-$1$ gauge fields.

Realising renormalisable Lagrangians within the fuzzy-sphere framework will allow for a systematic study of three-dimensional quantum field theories in the deep IR. This approach may address questions such as the IR fate of a given Lagrangian, and the duality between different field-theory Lagrangians~\cite{Dasgupta1981Duality,Seiberg2016Duality,Metlitski2015Duality}. For example, the free Majorana fermion that we have studied is conjectured to admit a dual Lagrangian description as the $\SO(N)_1$ Chern-Simons gauge theory coupled to a critical scalar in the $\SO(N)$ vector representation~\cite{Metlitski:2016dht,Aharony2016Duality,Cordova2017Duality}. A fuzzy sphere realisation of the latter would allow numerical tests of this duality which go beyond comparisons of the expected phase diagrams, such as those associated with fIQH-MQH transitions.

The correspondence with Lagrangians also provides an intuitive way to design models and understand the physical meaning of the tuning parameters on the fuzzy sphere. By assigning each Majorana fermion and each real scalar an `excitation' flavour in addition to a `reference' flavour, various field theories can be constructed in a similar way to the super-Ising theory, including
\begin{itemize}
    \item The $\rO(N_f)$ free Majorana fermion and Gross-Neveu-Yukawa theory~\cite{Gross1974,Erramilli:2022kgp}, 
    \item the $\cN=2$ Wess-Zumino model~\cite{Bobev:2015vsa} --- or equivalently the GNY theory of a Dirac fermion coupled to a complex scalar, and 
    \item the generalisation of Ginzburg-Landau description of a class of non-unitary minimal models to three space-time dimensions~\cite{Katsevich:2024jgq}.
\end{itemize}

Nevertheless, our current realisation of the super-Ising theory still requires the fine tuning of three relevant operators due to the absence of the parity symmetry or manifest supersymmetry. In the fuzzy-sphere models, parity is usually implemented as the particle-hole symmetry combined with certain flavour transformations. This is not directly applicable for models with microscopic bosons. Finding a way to implement the parity for the boson-fermion-mixed models would greatly simplify targeting the fermionic CFTs. Besides that, several further questions regarding the realisation of super-Ising theory deserve attention, \textit{e.~g.}, how the correspondence~\eqref{eq:op_suis} between the CFT and fuzzy-sphere operators holds numerically, whether it is possible to construct the emergent super-conformal generators $Q^\alpha$ and $S^\alpha$ in analogy with the conformal generators $P^\mu$ and $K^\mu$~\cite{Fardelli2024,Fan2024}, how the reference-state overlap and the small orbital-space entanglement hold for the super-Ising SCFT, \textit{etc.}

Instead of letting the supersymmetry emerge, one can also try to implement part of the fermionic charges of the super-conformal symmetry at the microscopic level. The idea of fuzzy super-sphere has indeed been explored over decades~\cite{Hasebe2004SUSY,Hasebe2005,Grosse1998}. The super-sphere $S^{2|2}$ is embedded in a super-space with three regular and two Grassmannian co-ordinates and is equipped with the $\OSp(1|2)$ super-rotation symmetry~\cite{Scheunert1977,Daumens1993}. We discuss this construction in more detail in Appendix~\ref{app:super_sphere}. The LLL on the super-sphere with a $4\pi q$ super-monopole carries a super-spin-$q$ representation, which branches into two $\SU(2)$ multiplets with spin-$q$ and $(q-\half)$ of opposite Grassmannian parity, in direct correspondence with our set-up. The local density-density interaction on the LLL also takes a similar form as the Hamiltonian~\eqref{eq:hmt} for the free Majorana fermion. However, the embedding of $\SU(2)$ rotations in $\Sp(4,\BR)$ conformal transformations does not lift to an embedding of the super-rotation group $\OSp(1|2)$ into the $\cN=1$ super-conformal group $\OSp(1|4)$. An embedding $\OSp(1|2) \subset U(1|2) \subset \OSp(2|4,\BR)$ into the $\cN=2$ superconformal group does exist, though. A more serious problem is that the super-rotation group $\OSp(1|2)$ does not admit any finite-dimensional non-trivial representation with a positive-definite inner product. If one is willing to relax the requirement of unitarity and allow for an indefinite inner product, this embedding may provide a route to engineer $\cN=2$ SCFTs: as the rotation group is enhanced to a conformal group in the IR, the $\OSp(1|2)$ fermionic super-rotation generators that mix the bosonic and fermionic orbitals must also become part of a larger symmetry super-group and $\OSp(2|4,\BR)$ is a natural candidate.

Besides supersymmetric theories, Chern-Simons-matter theories with fermionic content is also an interesting target. For example, the confinement transition of $\nu=1/3$ Laughlin state which describes its transition with a trivially gapped state, has $2\pi$-monopole with spin-$3/2$~\cite{Chester2017Monopole} which may be enhanced to the super-current in a $\cN=2$ supersymmetry, but the dual supersymmetric theory is expected to be gapped~\cite{Witten1999SUSY}. Another series of examples are the QED with Chern-Simons coupling~\cite{Lee2018QEDCS}, which describe transitions between Jain states~\cite{Jain1989Composite}. For example, the $N_f=3$ QED describes the transition between the states with fillings $\nu=1/3$ and $\nu=2/3$, and has $4\pi$-monopoles with spin-$1/2$ and $3/2$. The fuzzy sphere is a promising platform to study these theories, as the realisation of the proximate topological orders are natural with the fractional quantum Hall set-up, \textit{e.~g.}, the Laughlin states~\cite{Laughlin1983Anomalous} with filling $\nu=1/k$ realise the Abelian topological orders captured by Chern-Simons theories $\rU(1)_{-k}$, the Jain sequence~\cite{Jain1989Composite} with $\nu=p/(mp+1)$ realises the Abelian topological orders with composite fermion descriptions, and the Read-Rezayi sequence with $\nu=p/(mp+2)$~\cite{Read1998Rezayi}, as a natural extension of the Moore-Read states~\cite{Moore1991Pfaffian}, realises in general a series of non-Abelian topological orders. Another noticeable direction is to generalise the gauged Ising transition as $\SU(2)_1$ coupled to an adjoint fermion to other adjoint QCDs~\cite{Gomis2017Adjoint}, which may describe more exotic phase transitions.

Moving on to the Ising$^\ast$ transition, we explicitly demonstrate the physics of a topological defect of a CFT on the fuzzy sphere. Moreover, the gauging of the $\BZ_2$ symmetry offers opportunities to study conformal line defects in the Ising CFT. For example, the monodromy defect in 3d Ising CFT~\cite{Billo2013Monodromy} is attached to a half-infinite topological plane across which a $\BZ_2$ twist is inserted. After gauging the $\BZ_2$ symmetry, it becomes the vortex line of the $\BZ_2$ gauge field which is no longer attached to the topological plane. This means that it can be realised locally at antipodal points on the fuzzy sphere. The vortex line at the Ising$^\ast$ transition is closely related to the $\sigma$ anyon in the proximate bPf phase: once the bulk is gapped, the vortex line becomes the $\sigma$ worldline; under radial quantisation, this appears as two $\sigma$ anyons localised at the poles. The attempt to trap and separate anyons~\cite{Toke2007Trap} in the $\nu=5/2$ FQH state may give an hint on how to realise the vortex line in the Ising$^\ast$ CFT trap anyons in the bPf phase. Another interesting related question to explore is the self-dual confinement transition in the $\BZ_2$ topological order~\cite{Tupitsyn2008IsingGauge}: While condensing $e$ or $m$ alone in the $\BZ_2$ TO drives an Ising$^\ast$ type transition to the trivially gapped phase, the transition condensing the $e$ and $m$ anyon at the same time in a self-dual manner may fall in a different universality. Its nature attracts wide attention, and a few critical exponents have been computed in lattice models through Monte Carlo~\cite{Somoza2020SelfDual}. This question could be potentially addressed on the fuzzy sphere, as the $\BZ_2$ topological order could be produced by coupling the $\SU(2)_2$ (\textit{i.~e.}~bPf) to its opposite-chirality partner $\SU(2)_{-2}$ and condensing certain types of anyons. 

\acknowledgments

We would like to thank Jaume Gomis, Zixiang Hu, Ryan Lanzetta, Zlatko Papi\'c, Neville Joshua Rajappa, Subir Sachdev, Chong Wang, Ying-Hai Wu, Zechuan Zheng, Wei Zhu, and Yijian Zou for fruitful discussions. Z.~Z.~acknowledges support from the Natural Sciences and Engineering Research Council of Canada (NSERC) through Discovery Grants. Research at Perimeter Institute is supported in part by the Government of Canada through the Department of Innovation, Science and Industry Canada and by the Province of Ontario through the Ministry of Colleges and Universities. The numerical calculations are done using the package FuzzifiED~\cite{FuzzifiED}.

\paragraph*{Note added} Upon preparing this manuscript, we became aware of a parallel study~\cite{Voinea2025} that investigates the gauged Majorana CFT (or Majorana$^\ast$, which is bosonic) in a different set-up of quantum Hall bilayer transition. The Section~\ref{sec:superising} on the super-Ising SCFT was added during a revision on January 2026. It appears after the first realisation of supersymmetry on the fuzzy sphere~\cite{Tang2025} in the gauged super-Ising universality (or super-Ising$^\ast$, which is bosonic).

\appendix

\section{Edge CFT Character and Entanglement Spectrum Degeneracy Count}
\label{app:edge}

In this section, we give the detail on calculating the degeneracy count for the entanglement spectrum in Section~\ref{sec:phase} and Figure~\ref{fig:ent} from the character of the 2D chiral CFTs on the edge.

\paragraph{Fermionic integer quantum Hall phase (fIQH)} is described by $\rU(1)_1\cong\SO(2)_1$. The 2D chiral CFT on its edge is a pair of free chiral Majorana fermion that form a $\rU(1)$ symmetry and carry charge $Q=\pm 1$. The character of the edge CFT is
\begin{align}
    \chi^{(\text{fIQH})}(q,y)&=\prod_{Q=\pm 1}q^{1/48}\prod_{k\in\BZ+1/2}(1+y^Qq^k)\nonumber\\
    &=q^{1/24}(1+yq^{1/2}+y^{-1}q^{1/2}+y^2q^2+y^{-2}q^2+\dots)\nonumber\\
    &\qquad\qquad\times(1+q+2q^2+3q^3+5q^4+7q^5+11q^6+\dots)\nonumber\\
    &=q^{1/24}(1+q+2q^2+3q^3+5q^4+7q^5+11q^6+\dots\nonumber\\
    &\qquad\qquad+yq^{1/2}+yq^{3/2}+2yq^{5/2}+3yq^{7/2}+5yq^{9/2}+\dots\nonumber\\
    &\qquad\qquad+y^{-1}q^{1/2}+y^{-1}q^{3/2}+2y^{-1}q^{5/2}+3y^{-1}q^{7/2}+5y^{-1}q^{9/2}+\dots\nonumber\\
    &\qquad\qquad+y^2q^2+y^2q^3+2y^2q^4+3y^2q^5+5y^2q^6+\dots\nonumber\\
    &\qquad\qquad+y^{-2}q^2+y^{-2}q^3+2y^{-2}q^4+3y^{-2}q^5+5y^{-2}q^6+\dots\nonumber\\
    &\qquad\qquad+\dots)
    \label{eq:edge_fiqh}
\end{align}
Here $y=e^{2\pi iz}$ and $z$ is the fugacity that couples to the $\rU(1)$. Its chiral central charge is $c_-=1$. By state-operator correspondence, the coefficient in front of $y^Qq^{m-c_-/24}$ is the degeneracy in the entanglement spectrum with angular momentum $m$ in $z$-direction (compared with the ground state) and electric charge $Q$. The factorisation shows that the chiral edge modes in different charge sector have the same degeneracy but have the momentum $m$ shifted.

\paragraph{Majorana quantum Hall (MQH) with $\nu_K=3$} is described by $\SO(3)_1$. The 2D chiral CFT on its edge is three free chiral Majorana fermions. They form a vector under a $\SO(3)$ global symmetry on the edge, of which a $\rU(1)$ subgroup is manifest on the fuzzy sphere. Its character is
\begin{align}
    \chi^{(\text{MQH})}(q,y)&=\prod_{Q=0,\pm 1}q^{1/48}\prod_{k\in\BZ+1/2}(1+y^Qq^k)\nonumber\\
    &=q^{1/16}(1+yq^{1/2}+y^{-1}q^{1/2}+y^2q^2+y^{-2}q^2+\dots)\nonumber\\
    &\qquad\qquad\times(1+q^{1/2}+q+2q^{3/2}+3q^2+4q^{5/2}+5q^3+7q^{7/2}+\dots)\nonumber\\
    &=q^{1/16}(1+q^{1/2}+q+2q^{3/2}+3q^2+4q^{5/2}+5q^3+7q^{7/2}+\dots\nonumber\\
    &\qquad\qquad+yq^{1/2}+yq+yq^{3/2}+2yq^2+3yq^{5/2}+4yq^3+5yq^{7/2}+\dots\nonumber\\
    &\qquad\qquad+y^{-1}q^{1/2}+y^{-1}q+y^{-1}q^{3/2}+2y^{-1}q^2+3y^{-1}q^{5/2}+4y^{-1}q^3+\dots\nonumber\\
    &\qquad\qquad+y^2q^2+y^2q^{5/2}+y^2q^3+2q^{7/2}+3y^2q^4+4y^2q^{9/2}+5y^2q^5+\dots\nonumber\\
    &\qquad\qquad+y^{-2}q^2+y^{-2}q^{5/2}+y^{-2}q^3+2q^{7/2}+3y^{-2}q^4+4y^{-2}q^{9/2}+\dots\nonumber\\
    &\qquad\qquad+\dots)
    \label{eq:edge_mqh}
\end{align}

\paragraph{Bosonic Pfaffian (bPf)} is described by $\SU(2)_2$. As $\SU(2)$ is a double covering of $\SO(3)$, $\SO(3)_1\cong\SU(2)_2/\BZ_2$, \textit{i.~e.}, the $\SU(2)_2$ can be regarded as gauging the $\BZ_2$ fermion-parity symmetry from $\SO(3)_1$. Correspondingly, on its edge
\begin{align}
    \chi^{(\text{bPf})}(q,y)&=\half\left[\chi^{(\text{MQH})}(q,y)\pm(q^{1/2}\to-q^{1/2})\right]\nonumber\\
    &=(\text{$\chi^{(\text{MQH})}(q,y)$ with only $\BZ$ (or $\BZ+\half$) power of $q$})\nonumber\\
    &=q^{1/16}(q^{1/2}+2q^{3/2}+4q^{5/2}+7q^{7/2}+\dots\nonumber\\
    &\qquad\qquad+yq^{1/2}+yq^{3/2}+3yq^{5/2}+5yq^{7/2}+\dots\nonumber\\
    &\qquad\qquad+y^{-1}q^{1/2}+y^{-1}q^{3/2}+3y^{-1}q^{5/2}+5y^{-1}q^{7/2}\dots\nonumber\\
    &\qquad\qquad+y^2q^{5/2}+2q^{7/2}+4y^2q^{9/2}+\dots\nonumber\\
    &\qquad\qquad+y^{-2}q^{5/2}+2q^{7/2}+4y^{-2}q^{9/2}+\dots\nonumber\\
    &\qquad\qquad+\dots)\qquad\qquad\text{(taking $-$ and $\BZ+\half$ for $N_{mf}\in4\BZ+2$)}.
    \label{eq:edge_bpf}
\end{align}
The choice of $\pm$ depends on whether the ground state of the chiral CFT on the edge is shifted to an integer or half-integer $m$, \textit{i.~e.}~$+$ for $N_{mf}\in4\BZ$ and $-$ for $N_{mf}\in4\BZ+2$. Its chiral edge mode is the same as removing the $m\in\BZ+\half$ levels and keeping the $m\in\BZ$ part of the chiral edge mode in $\SO(3)_1$. 

\section{Lagrangian Description of the Gauged Ising Transition}
\label{app:u2cs}

In this section, we construct a Lagrangian that captures the transition between the bPf phase described by $\SU(2)_2$ and the MQH phase described by $\SU(2)_2/\BZ_2\cong\SO(3)_1$. We consider a $\rU(2)$ gauge field $\alpha+\tfrac{1}{2}a\,\BI_2$ where $\alpha$ is the traceless part and $a$ is the trace part.
\begin{multline}
    S=\int\frac{2}{4\pi}\tr\left(\alpha\,\rd\alpha+\tfrac{2}{3}\alpha^3\right)+\frac{1}{4\pi}a\,\rd a+\frac{2}{2\pi}b\,\rd a+\frac{1}{2\pi}A\,\rd a+\frac{1}{4\pi}A\,\rd A\\
    +\rd^3x\left[ \left((\partial_\mu-ia_\mu)\phi\right)^2+\frac{m^2}{2}\phi^2-\frac{\lambda}{4!}\phi^4 \right]
\end{multline}
We start with the
\begin{equation*}
    \rU(2)_{2,1}=\frac{\SU(2)_2\times\rU(1)_2}{\BZ_2}
\end{equation*}
Chern-Simons theory. The trace part $a$ is coupled to the electromagnetic field $A$. We couple $a$ to another dynamical $\rU(1)$ gauge field $b$ through a mutual Chern-Simons coupling. This coupling will Higgs $a$ from $\rU(1)$ to $\BZ_2$. This can be seen from
\begin{enumerate}
    \item The equation of motion of $b$ asserts that $\rd a=0$, and
    \item The insertion of a Wilson line of $b$ enforces an $a$-flux $\oint a\in\pi\BZ$ around it.
\end{enumerate}
This $\BZ_2$ gauge field is now the centre of the $\SU(2)$. We then couple the $\BZ_2$ gauge field $a$ to a real scalar $\phi$. The phase transition is controlled by the mass of the scalar.
\begin{enumerate}
    \item When $m^2>m_c^2$, the scalar field is gapped, and the $\BZ_2$ gauge field remains. The resulting theory is $(\SU(2)_2\times\BZ_2)/\BZ_2=\SU(2)_2$, which describes the topological order of bosonic Pfaffian phase.
    \item When $m^2<m_c^2$, the scalar field is condensed, this confines the $\BZ_2$ gauge field. The resulting theory is $\SU(2)_2/\BZ_2=\SO(3)_1$, which describes the $\nu_K=3$ MQH.
\end{enumerate}

\section{Realisation of Local Spinor Operators on the Fuzzy Sphere}
\label{app:corr}

The local Majorana fermion $\chi$ is a two-component spinor,
\begin{equation}
    \chi(\br)=(\{\chi^M(\br)\}_{M=\pm1/2})^\rT=\begin{pmatrix}
        \chi^{-1/2}(\br)\\\chi^{1/2}(\br)
    \end{pmatrix}.
\end{equation}
Here we denote the two components by an angular momentum index $M=\pm 1/2$, such that acting each component at the origin point will get us an eigenstate
\begin{align*}
    \begin{pmatrix}
        \chi^{-1/2}(0)\\0
    \end{pmatrix}|\BI\rangle&=|\chi_{l=1/2,m=-1/2}\rangle,&\begin{pmatrix}
        0\\\chi^{1/2}(0)
    \end{pmatrix}|\BI\rangle=|\chi_{l=1/2,m=1/2}\rangle.
\end{align*}

At each point on the sphere,
\begin{equation}
    \chi(\br)=\chi^{(+1/2)}(\br)\be_{(1/2)}+\chi^{(-1/2)}(\br)\be_{(-1/2)},
\end{equation}
where the two polarisations $\chi^{(s)}(\br)$ ($s=\pm 1/2$) of the spinor carry angular momentum $s$ under the rotation along an axis that passes $\br$. Their  polarisation vectors can be written as
\begin{equation}
    e_{(s)}^M=D^{1/2}_{sM}(0,\theta,\phi)=(-1)^{-s}Y^{(-s),1/2,M}(\br)
\end{equation}
Specifically,
\begin{align}
    \be_{(-1/2)}&=\begin{pmatrix}
        u\\-v
    \end{pmatrix},&\be_{(+1/2)}&=\begin{pmatrix}
        \bar{v}\\\bar{u}
    \end{pmatrix}.
\end{align}
Here $u=e^{-i\phi/2}\cos\theta/2$ and $v=e^{i\phi/2}\sin\theta/2$ compose the Hopf spinor. $(\be_{(-1/2)}(\br),\be_{(1/2)}(\br))$ is an orthonormal basis for spinor fields on the sphere. They satisfy
\begin{align}
    \bar{\be}^{(s)}\be_{(s')}&=\delta^s_{s'},&\bar{\be}^{(s)}&=(-1)^{-s}\be_{(-s)}^\rT\bg,&g_{mm'}&=\delta_{m+m',0}(-1)^{-m}.
\end{align}
Altogether, $(\be_{(-1/2)},\be_{(1/2)})$ composes the Wigner's $D$-matrix $D^l(\theta,\phi,\psi)$ with $l=1/2$ and $\psi=0$, which is a unitary matrix used to rotate spinning objects around the globe.

Here we explain some of the notations. The indices in the parentheses are the spin weight. The expression altogether must carry spin-weight $0$. We use a rule that the lower indices lower the angular momentum or spin weight and the upper indices raise them. The indices of spherical harmonics shall also be adapted to this rule. The indices are raised and lowered with a rule such that
\begin{align}
    u^m&=(-1)^{m}u_{-m},&u_m&=(-1)^{-m}u^{-m}.
\end{align}
The indices of the spherical harmonics follow the same logic, which makes it a bit different from the standard notation. This notation helps us keep track of all the conjugation relations and minus signs, \textit{e.~g.}, the conjugation of spherical harmonics is automatically taken care of
\begin{equation*}
    \Yb_{(s),l,m}=(-1)^{-s-m}Y^{(-s),l,-m}
\end{equation*}
as well as the orthonormal relations
\begin{equation*}
    e^{(s')}_{m}e^m_{(s)}=\delta^{s'}_s.
\end{equation*}
The polarisation can be decomposed into spherical components on the basis of spin-weighted spherical harmonics.
\begin{align}
    \chi^{(s)}(\br)&=\frac{1}{R^2}\sum_{l=1/2}^\infty\sum_{m=-l}^l\chi^{(s),l,m}Y^{(s),l,m}(\br)\nonumber\\
    \chi^{(s),l,m}&=\int\rd^2\br\,\Yb_{(s),l,m}(\br)\chi^{(s)}(\br).
\end{align}

Now we consider the two-point correlation functions of $\chi$. In the standard notation, the conformal kinematics requires
\begin{equation}
    \langle\chi^\alpha(x_1)\chi^\beta(x_2)\rangle=\frac{\gamma_\mu^{\alpha\beta}x_{12}^{\mu}}{(x_{12}^2)^{3/2}},
\end{equation}
where $x_{1,2}\in\BR^3$, $x_{12}=x_1-x_2$, $a,b$ are spinor indices, and $\gamma$ are the Pauli matrices. It can be equivalently written as
\begin{equation}
    \langle\chi^{m_1}(x_1)\chi^{m_2}(x_2)\rangle=\frac{\sqrt{2}C^{\half m_1; \half m_2}_{1,m}\hat{x}_{12}^m}{(x_{12}^2)^2}
\end{equation}
where
\begin{align}
    \hat{x}^m&=\sqrt{\frac{4\pi}{3}}Y^{1m}(\hat{x})=\left(\frac{1}{\sqrt{2}}e^{-i\phi}\sin\theta,\cos\theta,-\frac{1}{\sqrt{2}}e^{i\phi}\sin\theta\right)^\rT\nonumber\\
    &=C^{\half s_1;\half s_2}_{10}C^{1m}_{\half m_1;\half m_2}e_{(s_1)}^{m_1}(\hat{x})e_{(s_2)}^{m_2}(\hat{x}),
\end{align}
and $C^{l_1m_1;l_2m_2}_{lm}$ is the Clebsh-Gordan coefficient. We use this notation such that the contraction of indices are clearer.

We first consider the case where $x_2$ is at the origin point and $x_1$ is on the unit sphere. After Weyl transformation, it can be measured through the matrix element $\langle0|\chi^{(s)}(\br)|\chi^m\rangle$ on the fuzzy sphere.
\begin{equation}
    \langle0|\chi^{(s)}(\br)|\chi^m\rangle=\frac{1}{R}e^{(s)}_{m'}\langle\chi^m(\br)\chi^{m'}(0)\rangle=\frac{R^{-1}}{\sqrt{2\pi}}Y^{(s)\,\half m}(\br).
\end{equation}
Hence, for the spherical components,
\begin{equation}
    \langle 0|\chi^{(s)\,l'm'}|\chi^m\rangle=\frac{R^{-1}}{\sqrt{2\pi}}\delta_{l',\half}(-1)^m\delta_{m'+m,0}.
    \label{eq:chi_norm}
\end{equation}

We then consider the case where both points are on the unit sphere. After the Weyl transformation, it can be measured through $\langle 0|\chi^{(s_1)}(\br)\chi^{(s_2)}(\br)|0\rangle$ on the fuzzy sphere.
\begin{align}
    \langle 0|\chi^{(s_1)}(\br_1)\chi^{(s_2)}(\br_2)|0\rangle&=R^{-2}e^{(s_1)}_{m_1}(\br_1)e^{(s_2)}_{m_2}(\br_2)\langle\chi^{m_1}(x_1)\chi^{m_2}(x_2)\rangle\nonumber\\
    &=2i\delta_{s_1+s_2,0}x_{12}^{-2}\nonumber\\
    \langle 0|\chi_{(s_1)}(\br_1)\chi^{(s_2)}(\br_2)|0\rangle&=2(-1)^{1/2-s_1}\delta^{s_1}_{s_2}x_{12}^{-2}.
\end{align}
Hence, the correlation funciton of each spin-weighted component behaves exactly like a scalar correlator. In terms of the spherical components,
\begin{align}
    \langle 0|\chi_{(s)}\chi^{(s)}(\br_2)|0\rangle&=\sum_{l_1m_1l_2m_2}\langle\chi_{(s)\,l_1m_1}\chi^{(s)\,l_2m_2}\rangle\,\Yb_{(s)\,l_1m_1}(\br_1)Y^{(s)\,l_2m_2}\nonumber\\
    &=\sum_{l_1m_1l_2m_2}\langle\chi_{(s)\,l_10}\chi^{(s)\,l_10}\rangle\delta^{l_2m_2}_{l_1m_1}\,\Yb_{(s)\,l_1m_1}(\br_1)Y^{(s)\,l_2m_2}\nonumber\\
    &=\sum_l\langle\chi_{(s)\,l0}\chi^{(s)\,l0}\rangle\sum_m\Yb_{(s)\,lm}(\br_1)Y^{(s)\,lm}(\br_2)
\end{align}
where the second line comes from the rotation symmetry. To perform the summation, we express the spherical Harmonics in terms of the $D$-matrix
\begin{align}
    Y^{(s)\,lm}(\br_2)&=(-1)^s D^l_{-s,m}(0,\theta_2,\phi_2)\nonumber\\
    \bar{Y}_{(s)\,lm}(\br_1)&=(-1)^{-l}\Yb_{(-s)\,l,m}(-\br_1)\nonumber\\
    &=(-1)^{-l-s}\bar{D}_{sm}^{l}(0,\pi-\theta_1,\pi+\phi_1)\nonumber\\
    &=(-1)^{-l-s}D_{ms}^l(-\pi-\phi_1,-\pi+\theta_1,0)\nonumber\\
    \sum_m \bar{Y}_{(s)\,lm}(\br_1)Y^{(s)\,lm}(\br_2)&=(-1)^{-l}(D^l(0,\theta_2,\phi_2)D^l(-\pi-\phi_1,-\pi+\theta_1,0))_{-s,s}\nonumber\\
    &=(-1)^lD^l_{-s,s}(\psi',\pi-\theta_{12},\phi')\nonumber\\
    &=(-1)^le^{is(\phi'-\psi')}d_{-s,s}^l(\pi-\theta_{12})\nonumber\\
    &=e^{is(\phi'-\psi')}d_{-s,s}^l(\theta_{12})\nonumber\\
    &=e^{is(\phi'-\psi')}\frac{2l+1}{4\pi}\left(\sin\frac{\theta_{12}}{2}\right)^{2s}P_{l-s}^{(2s,0)}(\cos\theta_{12}).
\end{align}
In the last line, we only calculate for positive $s$, and negative $s$ follows similarly. There is an overall phase that depends on the relative position of the two points, so we focus on the amplitude. In the third last line, we have used the superposition of rotations, and $\theta_{12}$ is the angular distance between the two points. The $P_{n}^{(\alpha,0)}$ are the Jacobi polynomials. They are defined on the inteval $[-1,1]$ and has orthonormal relation
\begin{equation}
    \int\rd\Omega\,\sin^{2s}\frac{\theta}{2}P_{l-s}^{(2s,0)}(\cos\theta)P_{l'-s}^{(2s,0)}(\cos\theta)=\frac{4\pi}{2l+1}\delta_{ll'}.
\end{equation}
The components are
\begin{align}
    C_\chi(\theta_{12})&=\langle 0|\chi_{(1/2)}(\br_1)\chi^{(1/2)}(\br_2)|0\rangle=\sum_l C_{\chi,l}\frac{2l+1}{4\pi}\sin\frac{\theta_{12}}{2}P_{l-1/2}^{(1,0)}(\cos\theta_{12})\\
    C_{\chi,l}&=2\pi\int\sin\theta_{12}\rd\theta_{12}\,C_\chi(\theta_{12})\,\sin\frac{\theta_{12}}{2}P_{l-1/2}^{(1,0)}(\cos\theta_{12}).
\end{align}
For $C_\chi(\theta_{12})=(2\sin\theta_{12}/2)^{-2}$, the decomposed components are
\begin{equation*}
    C_{\chi,l}=2\pi.
\end{equation*}

We then briefly discuss how to decompose the local fermionic operators on the fuzzy sphere
\begin{align}
    \eta^{(1/2)}(\br)&=f^\dagger(\br)b(\br),&\eta^{(-1/2)}(\br)&=f^\dagger(\br)b(\br)
\end{align}
into spherical components:
\begin{align}
    \eta^{(1/2)}(\br)&=\eta^{(1/2)\,lm}Y^{(1/2)\,lm}(\br)\\
    \eta^{(1/2)\,lm}&=\int\rd^2\br\,\bar{Y}_{(1/2)\,lm}\eta^{(1/2)}(\br)\nonumber\\
    &=\int\rd^2\br\,\bar{Y}_{(1/2)\,lm}f^\dagger(\br)b(\br)\nonumber\\
    &=\sum_{m_1m_2}f^{\dagger\,m_1}b_{m_2}\int\rd^2\br\,\bar{Y}_{(1/2)\,lm}Y^{(q)\,qm_1}\bar{Y}_{(q-1/2),q-1/2,m_2}\nonumber\\
    &=\sum_{m_1m_2}f^{\dagger\,m_1}b_{m_2}\,\delta_{m_1,m_2+m}(-1)^{-q-m_1}\nonumber\\
    &\qquad\qquad\times\sqrt{\frac{2q(2q+1)(2l+1)}{4\pi}}\begin{pmatrix}
        l&q&q-\half\\-m&m_1&-m_2
    \end{pmatrix}\begin{pmatrix}
        l&q&q-\half\\1/2&-q&q-\half
    \end{pmatrix}.
\end{align}
The $\eta^{(-1/2)}$ follows similarly.

The $\eta^{(\pm1/2)}$ realises the two polarisations of $\chi$
\begin{equation}
    \eta^{(1/2)}=\alpha\chi^{(1/2)}(\br)+\dots
\end{equation}
The coefficient $\alpha$ can be settled through Eq.~\eqref{eq:chi_norm}
\begin{equation}
    \alpha=-iR\sqrt{2\pi}\langle 0|\eta^{(s),\half,-\half}|\chi^{\half}\rangle.
\end{equation}
Hence, 
\begin{equation}
    \langle 0|\chi_{(1/2)}(\br_1)\chi^{(1/2)}(\br_2)|0\rangle=R^{-2}\frac{\langle 0|\eta_{(1/2)}(\br_1)\eta^{(1/2)}(\br_2)|0\rangle}{\frac{1}{2\pi}|\langle 0|\eta^{(s),\half,-\half}|\chi^{\half}\rangle|^2}
\end{equation}
and the dimensionless correlator Eq.~\eqref{eq:corr_dl} is exactly expressed as 
\begin{equation}
    C_\chi^\ast(\theta_{12})=\frac{\langle 0|\eta_{(1/2)}(\br_1)\eta^{(1/2)}(\br_2)|0\rangle}{\frac{i}{\pi}|\langle 0|\eta^{(s),\half,-\half}|\chi^{\half}\rangle|^2}.
    \label{eq:corr_dl_exact}
\end{equation}

\section{The Fuzzy Super-Sphere}
\label{app:super_sphere}

\newcommand{\thb}{\bar{\theta}}

In this section we explain the idea of the fuzzy super-sphere, its connection with our set-up for free Majorana fermion, and the embedding of the super-rotation symmetry in the $\cN=2$ super-conformal group. 

\paragraph{Super-sphere and super-rotation} We consider the super-space $E(3|2)$ spanned by the regular co-ordinates $x_\mu$ and Grassmannian co-ordinates $\theta_\alpha$ where $\mu=1,2,3$ and the spinor index $\alpha=\pm$. The super-sphere is defined by the constraint $r^2=x^\mu x_\mu+\theta_\alpha\epsilon^{\alpha\beta}\theta_\beta$. It is invariant under the super-rotation symmetry $\OSp(1|2)$, generated by three bosonic generators $J_\mu$ and two fermionic generators $J_\alpha$~\cite{Daumens1993}
\begin{align}
    J_\mu&=L_\mu+\frac{1}{2}\theta_\alpha(\gamma_\mu)^\alpha{}_\beta\partial^\beta\nonumber\\
    V_\alpha&=\frac{i}{2}\left[(1+\half\thb\theta)\epsilon_{\alpha\beta}\partial^\beta-\theta_\beta (\gamma^\mu)^{\beta}{}_\alpha L_\mu\right]
    \label{eq:surot_gen}
\end{align}
where $\partial^\alpha=\partial/\partial\theta_\alpha$, $\bar{\theta}^\alpha=\epsilon^{\alpha\beta}\theta_\beta$ and $L_\mu$ is the usual orbital angular momentum $L_\mu=i\epsilon^{\mu\nu\rho}x_\nu\partial_\rho$. They have the following commutation relation:\footnote{The spinor indices are raised and lowered by $\epsilon^{\alpha\beta}$ and $\epsilon_{\alpha\beta}$ tensor. In the following we will omit the $\epsilon$.}
\begin{align}
    [J_\mu,J_\nu]&=i\epsilon_{\mu\nu\rho}J^\rho,&[J_\mu,V_\alpha]&=\half J_\beta(\gamma_\mu)^\beta{}_\alpha,&\{V_\alpha,V_\beta\}=\half J_\mu(\gamma^\mu)_{\alpha\beta}.
    \label{eq:alg_osp}
\end{align}

These generators are equipped with super-Hermitianity
\begin{align*}
    J_\mu^\ddagger&=J_\mu,&J_\alpha^\ddagger=\epsilon^{\alpha\beta}J_\beta,
\end{align*}
a notion enhancing complex conjugation in the super-space to the operator algebra. It has the properties
\begin{align}
    (A^\ddagger)^\ddagger&=(-1)^{\nu(A)}A,&(AB)^\ddagger&=(-1)^{\nu(A)\nu(B)}B^\ddagger A^\ddagger
    \label{eq:super_hermitian}
\end{align}
for arbitrary operators $A,B$ where $\nu(\bullet)=0,1$ is the Grassmannian parity. The super-Hermitian structure of the algebra cannot be realised as the adjoint in any finite-dimensional non-trivial representation with a positive-definite inner product.\footnote{The positivity requires $J_3$ to be positive-semi-definite
\begin{equation*}
    \langle\Psi|J_3|\Psi\rangle=\langle\Psi|\{V_+,V_-\}|\Psi\rangle=\|V_+|\Psi\rangle\|^2+\|V_-|\Psi\rangle\|^2\geq 0
\end{equation*}
which is only possible for states in the identity representation of its real part $\SU(2)$.} Nevertheless, the super-Hermitianity requires that the eigenvalues are either real or appear in complex-conjugate pairs, similar to the requirement of $\mathcal{PT}$ symmetry in Yang-Lee singularity. In the following, we relax the requirement of unitarity and allow for an indefinite inner product in order to construct a quantum-mechanical system on the super-sphere.

The non-unitary finite-dimensional representations of $\OSp(1|2)$ are labelled by super-spin-$j$ and a Grassmannian degree $\lambda\in\BZ_2$~\cite{Scheunert1977,Daumens1993}. The dimension of the super-spin-$j$ representation is $4j+1$. The components are labelled by angular momenta $l$ and magnetic number $m$. The quantum numbers take value as 
\begin{align*}
    j&\in\half\BZ,&l&=j\textrm{ or }j-\half,&m&=-l,-l+1,\dots,l.
\end{align*}
Each state $|j\lambda;lm\rangle$ has a Grassmannian number $\lambda+2(j-l)$, total super-spin $j$, total orbital angular momentum $l$, and the angular momentum $z$-direction $m$. Each representation branches into a two $\SU(2)$ multiplets with spin-$j$ and $(j-\half)$, and their Grassmannian parities are opposite. 

\paragraph{The fuzzy super-sphere} To build a model on the super-sphere, we need the basis of lowest Landau level. As on the ordinary sphere, this basis can be constructed from the Hopf spinor~\cite{Hasebe2005}. We begin by recalling how the spherical LLL arises from the Hopf spinor of $\CP^1$. The Hopf spinor $(u,v)^\rT$ defines a $\rU(1)$ bundle over $\CP^1$ whose Berry curvature corresponds to the monopole field on $S^2$. In Cartesian co-ordinates it may be written as
\begin{equation}
    \begin{pmatrix}
        u\\v
    \end{pmatrix}=\frac{1}{\sqrt{2r(r+x_3)}}\begin{pmatrix}
        r+x_3\\
        x_1+ix_2
    \end{pmatrix}.
\end{equation}
The spherical LLL is realised as the space of holomorphic sections of the appropriate power of this bundle, \textit{i.~e.},~as homogeneous polynomials in the Hopf spinor components
\begin{align}
    &&\phi_m(u,v)&=C_mu^{q+m}v^{q-m}&(m&=-q,\dots,q)
\end{align}
where $C_m$ is a normalisation constant. In complete analogy, the Hopf spinor of $\CP^{1|1}$ contains two bosonic and one fermionic component
\begin{equation}
    \begin{pmatrix}
        u\\v\\\xi
    \end{pmatrix}=\frac{1}{\sqrt{2r^3(r+x_3)}}\begin{pmatrix}
        (r+x_3)\left(r-\frac{1}{4(r+x_3)}\thb\theta\right)\\
        (x_1+ix_2)\left(r+\frac{1}{4(r+x_3)}\thb\theta\right)\\
        -(r+x_3)\theta_+-(x_1+ix_2)\theta_-
    \end{pmatrix}.
\end{equation}
It yields the super-spherical LLL
\begin{align}
    &&\phi_m(u,v,\xi)&=C_mu^{q+m}v^{q-m}&(m&=-q,\dots,q)\nonumber\\
    &&\psi_m(u,v,\xi)&=C'_mu^{q-\half+m}v^{q-\half-m}\xi&(m&=-q+\half,\dots,q-\half).
    \label{eq:susy_lll}
\end{align}
It carries a super-spin-$q$ representation of the $\OSp(1|2)$ super-rotation group. It decomposes into $2q+1$ Grassmannian-even orbitals and $2q$ Grassmannian-odd orbitals with respective orbital angular momenta $q$ and $q-\half$. By intepreting the Grassmannian-even orbitals as fermions and the Grassmannian-odd orbitals as bosons, the structure is in complete correspondence with our set-up for the Majorana fermion. The mismatch $N_{mf}-N_{mb}=1$ in the set-up for the Majorana fermion is inherited from the $1/2$ angular-momentum difference in the two $\SU(2)$ multiplets. A crucial difference, however, is that implementing the full super-rotation symmetry requires a non-unitary (indefinite) realisation: as discussed above, the super-Hermitian structure of the algebra does not admit a non-trivial positive-definite finite-dimensional representation.

The physical set-up for a fuzzy super-sphere involves a super-monopole that exerts a connection on the super-sphere~\cite{Grosse1998,Hasebe2005}
\begin{align}
    A_\mu&=-\frac{q}{r(r+x_3)}\epsilon_{\mu\nu3}x^\nu\left[ 1+\frac{2r+x_3}{4r^2(r+x_3)}\thb\theta\right]\nonumber\\
    A_\alpha&=\frac{iq}{r^3}x_\mu\theta_\beta(\gamma^\mu)^\beta{}_\alpha.
\end{align}
The single-particle Hamiltonian becomes
\begin{equation}
    H_0=\frac{1}{2mr^2}(D^\mu D_\mu+D_\alpha\epsilon^{\alpha\beta}D_\beta)
\end{equation}
where the covariant operators
\begin{align*}
    D_\mu=&=\partial_\mu+iA_\mu,&D_\alpha&=\partial_\alpha-iA_\alpha.
\end{align*}

The single-particle states are the monopole super-spherical harmonics $Y^{(q)}_{jlm}$
\begin{align}
    j-s&\in\BZ&Y^{(q)}_{jjm}&=\frac{1-\half j\thb\theta}{\sqrt{2j+1}}Y^{(q)}_{jm}\nonumber\\
    &&Y^{(q)}_{j,j-\half,m}&=i\left[\sqrt{\frac{j+m+\half}{2j+1}}\theta_-Y^{(q)}_{j,m+\half}-\sqrt{\frac{j-m+\half}{2j+1}}\theta_+Y^{(q)}_{j,m-\half}\right]\nonumber\\
    j-s&\in\BZ+\half&Y^{(q)}_{j,j-\half,m}&=\frac{1-\half(j+\half)\thb\theta}{\sqrt{2j}}Y^{(q)}_{j-\half,m}\nonumber\\
    &&Y^{(q)}_{jjm}&=i\left[\sqrt{\frac{j-m}{2j}}\theta_-Y^{(q)}_{j-\half,m+\half}+\sqrt{\frac{j+m}{2j}}\theta_+Y^{(q)}_{j-\half,m-\half}\right].
\end{align}
Here $j$ is the total super-spin index, $l$ is the total orbital angular momentum index, and the indices takes value as
\begin{align*}
    j&=q,q+\half,q+1,\dots,&l&=j,j-\half,&m&=-l,-l+1,\dots,l.
\end{align*}
and the lowest Landau level wave-functions are 
\begin{align*}
    \phi_m&=Y_{qqm}^{(q)}(x_\mu,\theta_\mu),&\psi_m&=Y_{q,q-\half,m}^{(q)}(x_\mu,\theta_\mu)
\end{align*}

We consider the super-field operator on the super-sphere 
\begin{equation}
    \Upsilon(x_\mu,\theta_\alpha)=\Psi_0(x_\mu)+\bar{\theta}^\alpha\Phi_\alpha(x_\mu)+\half\thb\theta\Psi_1(x_\mu)
\end{equation}
where $\Psi_{0,1}$ are fermionic fields and $\Phi_\alpha$ are bosonic fields. We can then project it onto the lowest Landau level~\eqref{eq:susy_lll}
\begin{equation}
    \Upsilon(x_\mu,\theta_\alpha)=\sum_mf^\dagger_mY_{qqm}^{(q)}(x_\mu,\theta_\mu)+\sum_mb^\dagger_mY_{q,q-\half,m}^{(q)}(x_\mu,\theta_\mu).
\end{equation}
Here $b^\dagger_m$ and $f^\dagger_m$ are the creation operators on the lowest Landau level. 

Like on the fuzzy sphere, we can express the interaction Hamiltonian in terms of $f^{(\dagger)}_m$ and $b^{(\dagger)}_m$ on the fuzzy super-sphere. The interaction theories may be built as transitions between gapped states like the super-Haldane-Laughlin states~\cite{Hasebe2004SUSY,Hasebe:2005cm,Hasebe:2006jf,Hasebe:2007au,Faizal:2011cd} analogous to the Laughlin states on the regular LLL. A convenient approach to build the interaction Hamiltonian is through the density operator
\begin{equation*}
    n(x_\mu,\theta_\alpha)=\Upsilon^\dagger(x_\mu,\theta_\alpha)\Upsilon(x_\mu,\theta_\alpha)
\end{equation*} 
As an example, the local density-density interaction of a single LLL reads
\begin{multline}
    \int\rd^2x_\mu\,\rd^2\theta_\alpha\,n(x_\mu,\theta_\alpha)^2\\=\sum_{\{m_i\}}\Big[\alpha^{(b)}M_{m_1m_2m_3m_4}^{(b)}b^\dagger_{m_1}b^\dagger_{m_2}b_{m_3}b_{m_4}+\alpha^{(bf)}M_{m_1m_2m_3m_4}^{(bf)}b^\dagger_{m_1}f^\dagger_{m_2}f_{m_3}b_{m_4}\\
    +\alpha^{(t)}M_{m_1m_2m_3m_4}^{(t)}(b^\dagger_{m_1}b^\dagger_{m_2}f_{m_3}f_{m_4}-f^\dagger_{m_4}f^\dagger_{m_3}b_{m_2}b_{m_1})\Big]
\end{multline}
where the co-efficients $\alpha^{(b)},\alpha^{(bf)}$ and $\alpha^{(t)}$ are constrained by the super-rotation. It has similar form with the Majorana fermion Hamiltonian~\eqref{eq:hmt} with $\mu=0$, except that the super-Hermitianity~\eqref{eq:super_hermitian} requires $b^\dagger_{m_1}b^\dagger_{m_2}f_{m_3}f_{m_4}$ and $f^\dagger_{m_4}f^\dagger_{m_3}b_{m_2}b_{m_1}$ to have opposite co-efficients (instead of the same under Hermitianity). This interaction alone produces the super-Haldane-Laughlin state.

\paragraph{Embedding into the super-conformal algebra} We need to embed the super-rotation group $\OSp(1|2)$ into the super-conformal algebra. The super-conformal generators in $d=3$ generates a super-Lie group $\OSp(\cN|4)$ with bosonic dimension $10+\half\cN(\cN-1)$ and fermionic dimension $4\cN$~\cite{Park2000,Gates2001}. The bosonic generators are $P_\mu,M_{\mu\nu},D,K_\mu,A_{ab}$ and fermionic generators are $Q^{a\alpha}, S^{a\alpha}$ where $a=1,\dots,\cN$ is the super-charge index. We first consider the case of $\cN=1$. Although $\OSp(1|2)$ is indeed a sub-group of $\OSp(1|4)$, the embedding is in general impossible due to the incompatiblity of the commutators. Besides the super-rotation, one also needs to embed the time translation $-iH$ that commutes with every other generator. The commutatition with rotation $[H,M^{\mu\nu}]=0$ constrains that $-iH=D$, but no combination of the fermionic generators $Q$ and $S$ commutes with $D$. Hence, there exists no consistent embedding of super-rotation $\OSp(1|2)$ and $-iH$ into the super-conformal group $\OSp(1|4)$

We then consider a larger super-conformal group $\OSp(2|4)$ for $\cN=2$. Besides the conformal generators, the generators include the fermionic $Q_\alpha,\bar{Q}_\alpha,S_\alpha,\bar{S}_\alpha$ and an additional bosonic $\rU(1)_R$ generator $A$. We construct the embedding
\begin{align}
    J_\mu&=\epsilon_{\mu\nu\rho}M^{\nu\rho}\nonumber\\
    V_\alpha&=\half(Q_\alpha+i\bar{S}_\alpha).
\end{align}
One can check explicitly that they satify the super-rotation algebra $\osp(1|2)$~\eqref{eq:alg_osp}. Here we list the relevant non-zero commutators
\begin{align}
    [M_{\mu\nu},Q_{\alpha}]&=i\half(\gamma_{[\mu}\gamma_{\nu]})_{\alpha\beta} Q^{\beta}\nonumber\\
    [M_{\mu\nu},\bar{S}_{\alpha}]&=i\half(\gamma_{[\mu}\gamma_{\nu]})_{\alpha\beta} \bar{S}^{\beta}\nonumber\\
    \{Q_{\alpha},\bar{S}_{\beta}\}&=-i(\gamma^{[\mu}\gamma^{\nu]})_{\alpha\beta} M_{\mu\nu}+2i\epsilon_{\alpha\beta} (D-A).
\end{align}

The time translation generated by $-iH$ cannot be identified as the dilatation $D$ anymore, as it does not commute with the fermionic generators 
\begin{equation*}
    [D,V_\alpha]=\tfrac{i}{4}(Q_\alpha+i\bar{S}_\alpha),
\end{equation*}
but $V_\alpha$ commutes with the combination of $D$ and the generator of the $\rU(1)_R$ symmetry $A$, which we can now define as Hamiltonian.
\begin{align}
    -iH&=D+\half A,&[H,V_\alpha]&=0.
\end{align}
Its eigenvalue $E$ would be a combination of the scaling dimension $\Delta$ and the $U(1)_R$ charge $r$, where we denote 
\begin{equation*}
    E=\Delta+\half r
\end{equation*}
as the energy. In this way, we have embedded all the manifest symmetries into the super-conformal group.

We comment on the rest of the super-conformal generators, which include 
\begin{align*}
    U_\alpha&=\half(Q_\alpha-i\bar{S}_\alpha),&&\bar{Q},&&S,&&\text{and}&-i\bar{H}&=D-\half A.
\end{align*}
These generators have commutation relations with the Hamiltonian
\begin{align}
    [H,U^\alpha]&=0,&[H,\bar{Q}^\alpha]&=\bar{Q}^\alpha,&[H,S^\alpha]&=-S^\alpha.
\end{align}
Therefore, $U$ does not change the energy, $\bar{Q}$ increases the energy by $1$ and $S$ decreases the energy by $1$. 

\paragraph{Super-conformal multiplet structure} We then consider how the energy levels organise themselves into multiplets in our set-up. The conformal multiplets combine into super-conformal multiplets~\cite{Cordova2019}. Each multiplet is marked by $(l,\Delta,r)$ of its primary, containing the Lorentz spin $l$, scaling dimension $\Delta$ and $U(1)_R$ charge $r$. To be more general, we list the super-conformal descendants in a long multiplet.
\begin{equation}
    \xymatrix{
        *+[F]{E}&(l,\Delta,r) \ar[r]^{Q}\ar[d]^{\bar{Q}}&(l\pm\half,\Delta+\half,r-1) \ar[r]^Q \ar[d]^{\bar{Q}}& (l,\Delta+1,r-2)\ar[d]^{\bar{Q}}\\
        *+[F]{E+1}&(l\pm\half,\Delta+\half,r+1) \ar[r]^{Q}\ar[d]^{\bar{Q}}&(l\oplus l\oplus l\pm1,\Delta+1,r) \ar[r]^Q \ar[d]^{\bar{Q}}& (l\pm\half,\Delta+\halves{3},r-1)\ar[d]^{\bar{Q}}\\
        *+[F]{E+2}&(l,\Delta+1,r+2) \ar[r]^{Q}&(l\pm\half,\Delta+\halves{3},r+1) \ar[r]^Q & (l,\Delta+2,r)
    }
    \label{eq:multp}
\end{equation}
Here we organise the table such that each line have the same eigenvalue under $H$, and for the primary $E=\Delta+\half r$. After considering the super-rotation symmetry, two multiplets whose conformal primaries have the same energy $E$ and spin $l=j$ and $l=j-\half$ combine into one, which we denote as `bi-multiplet.' Each bi-multiplet is marked by $[j,E]$ containing the super-spin $j$ and energy $E$. Subsequently, considering the first line of Diagram~\eqref{eq:multp}, they combine into two bi-multiplets
\begin{equation}
    \left\{\begin{array}{c}
        (l,\Delta,r)\\(l\pm\half,\Delta+\half,r-1)\\(l,\Delta+1,r-2)
    \end{array}\right\}\longrightarrow\left\{\begin{array}{c}
        [l,E]\\{[l-\half,E]}
    \end{array}\right\}
\end{equation}
where $E=\Delta+\half$. These two bi-multiplets are connected by the spinor generator $U$. To show the combination explicitly, we look at the simplest example of a scalar super-conformal primary $\Phi$. The conformal multiplets of $\Phi,Q\Phi,Q^2\Phi$ will be re-organised into two bi-multiplets $[\half]$ and $[0]$, where $Q^2=Q^\alpha\epsilon_{\alpha\beta}Q^\beta$ 
\begin{align}
    [\half,E]&=(Q\Phi)\oplus(Q^2\Phi-2i(\Delta+r)\Phi)\nonumber\\
    [0,E]&=(Q^2\Phi+2i(\Delta+r)\Phi).
\end{align}
The super-conformal primary does not belong to any of the two bi-multiplets. Instead, the content of the bi-multiplets are linear combinations of the primary and its super-conformal descendants.

Therefore, the 16 conformal multiplets in the super-conformal multiplet can be re-organised into 8 bi-multiplets.
\begin{equation}
    \xymatrix{
        *+[F]{E}& [j,E] \ar@{<->}[r]^{U} \ar[d]^{\bar{Q}} & [j+\half,E] \ar[d]^{\bar{Q}} \\
        *+[F]{E+1}& [j\pm\half,E+1] \ar@{<->}[r]^{U} \ar[d]^{\bar{Q}} & [j\oplus j+1,E+1] \ar[d]^{\bar{Q}} \\
        *+[F]{E+2}& [j,E+2] \ar@{<->}[r]^{U} & [j+\half,E+2]
    }
\end{equation}
Notably, one hallmark of this super-conformal structure is that each level is two-fold degenerate. These super-conformal tower, together with the conformal descendants, determine the integer-level structure in super-conformal symmetry.

Special attention should be paid to the short super-current super-conformal multiplets, which consists all the conserved currents of the super-conformal symmetry, including 
\begin{enumerate}
    \item Stress tensor $T^{\mu\nu}$ with quantum numbers $l=2,\Delta=3,r=0$, which integrates into the conformal charges $P_\mu,D,K_\mu$ and $M_{\mu\nu}$;
    \item Supersymmetry current $G^{\mu\alpha}$ with quantum numbers $l=\halves{3},\Delta=\halves{5},r=1$ which integrates into the super-charges $Q^\alpha,\bar{Q}^\alpha$, and its conjugate $\bar{G}^{\mu\alpha}$ with $l=\halves{3},\Delta=\halves{5},r=-1$;
    \item The conserved $\rU(1)_R$ current with quantum numbers $l=1,\Delta=2,r=0$.
\end{enumerate}
These four conserved currents are connected by the super-charges $Q$ and $\bar{Q}$, and they combine into two bi-multiplets on the super-sphere. The $J^\mu_R$ and $\bar{G}^{\mu\alpha}$ combine into a bi-multiplet with $j=\halves{3}$ and $E=2$; the $G^{\mu\alpha}$ and $T^{\mu\nu}$ combine into a bi-multiplet with $j=2$ and $E=3$.
\begin{equation}
    \xymatrix{
        & (l,\Delta,r) & [j=\halves{3},E=2] \ar@{=}[ld] \ar@{=}[rd] & (l,\Delta,r) \\
        *+[F]{E=2} & J_R^\mu (1,2,0) \ar[rr]^Q \ar[dd]^{\bar{Q}} & & \bar{G}^{\mu\alpha} (\halves{3},\halves{5},-1)\ar[dd]^{\bar{Q}}\\
        & & [j=2,E=3] \ar@{=}[ld] \ar@{=}[rd] & \\
        *+[F]{E=3} & G^{\mu\alpha}(\halves{3},\halves{5},1) \ar[rr]^Q & & T^{\mu\nu} (2,3,0)\\
    }
    \label{eq:str_ten}
\end{equation}

\section{The Higher Spectrum of the Super-Ising SCFT}
\label{app:suis_full}

\begin{figure}[htbp]
    \centering
    \includegraphics[width=\linewidth]{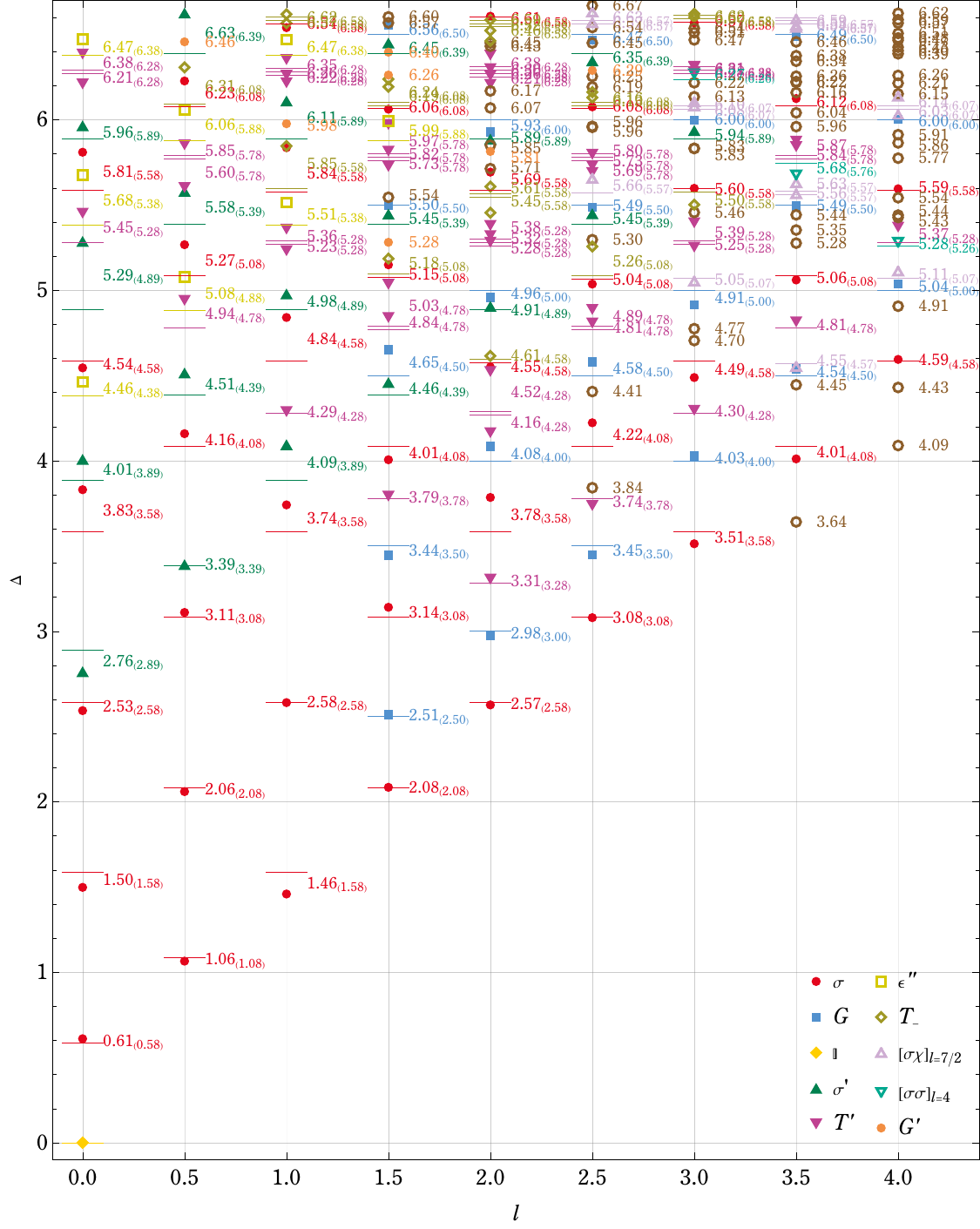}
    \caption{The spectrum of super-Ising field theory up to $\Delta\approx 6$ and $l=4$ at $N_{mf}=9$. The symbols marks the measured scaling dimensions and the bars marks the expected values from conformal bootstrap. Each identified state is labelled by the measured and expected (in parentheses in subscript) scaling dimensions. In the calculation, we fix $t=1.5,U=0.25,g=1$ and take $h,\mu_1,\mu_b$ as in Table~\ref{tbl:para_suis}.}
    \label{fig:suis_spec_full}
\end{figure}

In the ED calculation, we calculate the 120 lowest states respectively in the $L^z=0$ and $1/2$ sectors. These cover the spectrum up to $\Delta\approx 6$. We plot all these states in Figure~\ref{fig:suis_spec_full}. All the states up to $l\leq 2,\Delta\leq 5.5$, and most states with $l\leq 3$ can be identified as super-conformal states. 

The identified multiplets from bootstrap result~\cite{Atanasov2022SuperIsing} below $\Delta=6$ include\footnote{The space-time parity is not accessible in our calculation.}
\begin{enumerate}
    \item The elementary fields $\sigma$ with $l=0$, $\Delta_\sigma=0.5844435(83)$ and parity-odd,
    \item The super-current $G$ with $l=3/2$ and $\Delta_G=2.5$,
    \item $\sigma'$ corresponding to the fermion mass $\bar{\chi}\chi$ with $l=0$, $\Delta_\sigma=2.8869(25)$ and parity-odd,
    \item $T'^{\mu\nu}$ in the twisted trajectory of $[\sigma\sigma]$ identified through the extremal spectrum with $l=2$, $\Delta_{T'}=3.28(1)$ and parity-even,
    \item $\epsilon''$ identified through the extremal spectrum with $l=2$, $\Delta_{\epsilon''}=4.38(1)$ and parity-even,
    \item $T_-^{\mu\nu}$ in the twisted trajectory of $[\sigma\epsilon]$ identified through the extremal spectrum with $l=2$, $\Delta_{T_-}=4.58(1)$ and parity-odd,
    \item $[\sigma\chi]_{l=7/2}$ in the twisted trajectory of $[\sigma\chi]$ with $l=7/2$ and $\Delta_{[\sigma\chi]_{l=7/2}}\approx4.570$, and
    \item $[\sigma\sigma]_{l=4}$ in the twisted trajectory of $[\sigma\sigma]$ with $l=4$, $\Delta_{[\sigma\sigma]_{l=4}}\approx5.257$.
    \item $G'^{\mu\alpha}$ at $l=3/2$ and $\Delta\approx5.28$ that is observed in the extremal spectrum ($5.78$ in the $Q^+\mathcal{F}_-^{3/2}$ sector) by the conformal bootstrap, but is subject to large numerical uncertainty.
\end{enumerate}


\providecommand{\href}[2]{#2}\begingroup\raggedright\endgroup

\end{document}